\documentstyle[11pt]{article}

\newcommand{\mbf}[1]{\mbox{\boldmath ${#1}$}}
\newcommand{\vx}{{\bf x}}

\newcommand{\vz}{{\bf z}}
\newcommand{\vn}{{\bf n}}
\newcommand{\ve}{{\bf e}}
\newcommand{\vE}{{\bf E}}
\newcommand{\vmu}{\underline{{\mbf \mu}}}

\newcommand{\Ref}[1]{(\ref{#1})}
\newcommand{\binom}[2]{\left(\!\!\bma{c} {#1}\\ {#2}\ema\!\!\right)}

\newcommand{\beps}{\mbf{\eps} }
\renewcommand{\b}{{\hat \rho}}
\newcommand{\U}{{\rm U}}
\newcommand{\sgn}{{\rm sgn}}
\newcommand{\eps}{\varepsilon}
\newcommand{\xx}{\stackrel {\scriptscriptstyle \times}{\scriptscriptstyle \times}}
\newcommand{\xxa}{\stackrel {\scriptscriptstyle \times}{\scriptscriptstyle \times} \!}
\newcommand{\xxe}{\! \stackrel {\scriptscriptstyle \times}{\scriptscriptstyle \times}}
\newcommand{\PiL}{\mbox{$\frac{\pi}{L}$}} 
\newcommand{\tPiL}{\mbox{$\frac{2\pi}{L}$}} 
\newcommand{\half}{\mbox{$\frac{1}{2}$}}

\newfam\msbfam
\batchmode\font\twelvemsb=msbm10 scaled\magstep1 \errorstopmode
\ifx\twelvemsb\nullfont\def\Bbb{\bf}
\else	\catcode`\@=11
	\font\tenmsb=msbm10 \font\sevenmsb=msbm7 \font\fivemsb=msbm5
	\textfont\msbfam=\tenmsb
	\scriptfont\msbfam=\sevenmsb \scriptscriptfont\msbfam=\fivemsb
	\def\Bbb{\relax\ifmmode\expandafter\Bbb@\else
 		\expandafter\nonmatherr@\expandafter\Bbb\fi}
	\def\Bbb@#1{{\Bbb@@{#1}}}
	\def\Bbb@@#1{\fam\msbfam\relax#1}
	\catcode`\@=\active
\fi
\newcommand{\R}{{\Bbb R}}
\newcommand{\C}{{\Bbb C}}
\newcommand{\Z}{{\Bbb Z}}
\newcommand{\N}{{\Bbb N}}

\newcommand{\cC}{{\cal C}}
\newcommand{\cD}{{\cal D}}
\newcommand{\cO}{{\cal O}}
\newcommand{\cP}{{\cal P}}
\newcommand{\cH}{{\cal H}}
\newcommand{\cF}{{\cal F}}
\newcommand{\cE}{{\cal E}}
\newcommand{\cG}{{\cal G}}
\newcommand{\cS}{{\cal S}}
\newcommand{\cW}{{\cal W}}
\newcommand{\cV}{{\cal V}}
\newcommand{\cJ}{{\cal J}}

\newcommand{\QED}{\hfill$\Box $}
\newcommand{\eq}{\begin{equation}}
\newcommand{\eqend}{\end{equation}}
\newcommand{\eqa}{\begin{eqnarray}}
\newcommand{\nonueqa}{\begin{eqnarray*}}
\newcommand{\eqaend}{\end{eqnarray}}
\newcommand{\nonueqaend}{\end{eqnarray*}}
\newcommand{\nonu}{\nonumber \\ \nopagebreak}
\newcommand{\bma}[1]{\begin{array}{#1}}
\newcommand{\ema}{\end{array}}
\newcommand{\bc}{\begin{center}}
\newcommand{\ec}{\end{center}}

\begin{document}
\begin{flushright}
May 4, 1998
\end{flushright}
\vspace{.4cm}

\begin{center}

{\Large \bf Loop groups, anyons and the Calogero-Sutherland model}\\
\vspace{1 cm}
{\large Alan L. Carey$^a$ and Edwin Langmann$^b$}\\
\vspace{0.3 cm}
{\em $^a$ Department of Pure Mathematics, University of Adelaide\\ 
$^b$Theoretical Physics, Royal Institute of Technology, S-10044 Sweden}\\
\end{center}

\begin{abstract}
The positive energy representations of the loop group of $\U(1)$ are 
used to construct a boson-anyon correspondence.  We compute all the 
correlation functions of our anyon fields and study an anyonic 
$W$-algebra of unbounded operators with a common dense domain.  This 
algebra contains an operator with peculiar exchange relations with the 
anyon fields.  This operator can be interpreted as a second quantised 
Calogero-Sutherland (CS) Hamiltonian and may be used to solve the 
CS model.  In particular, we inductively construct all eigenfunctions 
of the CS model from anyon correlation functions, for all particle 
numbers and positive couplings.
\end{abstract}

\section{Introduction}

The viewpoint of Graeme Segal \cite{PS}, \cite{SegWil} on integrable 
systems links the infinite dimensional Grassmanian approach of Sato 
\cite{Sato} with the representation theory of loop groups.  These two 
points of view overlap in the study of two dimensional quantum field 
theories.  In the Sato approach, as in much of the physics literature, 
quantum field theory is regarded as an algebraic theory in which the 
usual Hilbert space formalism is absent.  The Segal approach on the 
other hand deals with positive energy representations of loop groups 
in Hilbert spaces.  Reconciling these points of view can be quite 
difficult although this has been done for many cases (see for example 
\cite{CR,CHMS,BMT}).  One way of thinking about the Segal approach is that 
it revolves around a Hilbert space definition of vertex operators.  
The algebraic approach to vertex operators is much studied in 
connection with Kac-Moody algebras \cite{Kac, F} and may be regarded as 
the Lie algebraic version of the loop group projective representation 
theory.  These Segal vertex operators arise from a boson field theory 
and were previously studied in a formal way in \cite{Skyrme,C,M} and 
made more precise in \cite{StrWil,DFZ}).  In this approach one 
regularises the vertex operators so that they are proportional to 
operators representing loop group elements and then, after taking an 
appropriate limit, one finds that they generate fermions in some cases 
(the boson-fermion correspondence) and operators forming a Kac-Moody 
algebra in others \cite{PS,Seg,CR,CH}, depending on the precise form of 
the cocycle in the loop group projective representation.

We may summarize the present paper as enlarging the loop group 
representation theory to encompass a boson-anyon correspondence.  Our 
results extend those of the previous paragraph in that we construct, 
from a certain positive energy loop group representation, Segal-type 
vertex operators on a Hilbert space which have, as their limits, anyon 
field operators.  These anyon field operators applied to the vacuum, 
or cyclic vector, give new vectors in the Hilbert space which can be 
interpreted as anyon states.  Each $N$-particle anyon sector carries a 
representation of the braid group.  The construction builds in 
fractional statistics from the outset, the precise statistics 
depending on the choice of anyon vertex operator.

The idea of using a vertex operator construction to obtain particles 
with anyon type statistics is not new, see for example \cite{Kl} 
and more recently \cite{AMOS1,AMOS2,Iso,H,MS} and references therein.  
However the vertex operators described in these more recent references 
are not defined on the Fermion Fock space as limits of implementors of 
fermion gauge transformations.  In other words they do not come from 
loop group elements.  Indeed it is difficult to give a precise meaning 
to them at all and we do not attempt to do so here.  Our vertex 
operators can be seen to have similar formal properties to those 
appearing in the papers mentioned, but are well defined in terms of 
positive energy representations of loop groups in the sense of 
\cite{PS}.

The benefits of our approach are the following.  First there is a 
quantum Hamiltonian acting on the anyon states.  This we believe 
resolves a long standing difficulty in the study of anyons in that it 
provides a basis for models incorporating interactions.  Second we 
obtain a unifying view of a number of interesting ideas that have 
emerged in recent times in the physics literature.  The most important 
of these is the connection with the Calogero-Sutherland (CS) model 
\cite{AMOS1,AMOS2,Iso,MS} (see also \cite{H,HLV,BHV,Poly}).  
Specifically we find that $n$-point anyon correlation functions 
provide useful building blocks for solutions to the CS system.  
Comparing with the known solutions of the CS system \cite{Fo2} we find 
that Jack polynomials \cite{St} may be expressed in terms of anyon 
correlation functions.  (Similar relations were previously obtained by 
different methods in \cite{Fo1}.)

{}From this point of view the anyon Hamiltonian is a second quantized 
CS Hamiltonian.  The final connection we make is with $W$-algebras, 
again a connection which has been known from other approaches for some 
time \cite{AMOS1,AMOS2,Iso,MS}.  In this paper we do not recover the 
full import of the $W$-algebra connection in the anyon case.  This is 
a matter we intend to develop more fully elsewhere.  However we do 
construct that part of the $W$-algebra that we need as an algebra of 
unbounded operators with a common dense domain.  This suffices for our 
purposes, namely the construction of an anyon Hamiltonian, 
constructing the CS model solutions as anyon 
correlation functions, obtaining the link with Jack polynomials, and 
finding the algebraic relations of the Hamiltonian with the anyon 
fields.

\section{Summary}

This paper contains a number of technical sections.  In order to make 
the results accessible we present a summary here.  At the same time we 
take the opportunity to introduce some of our notation.  However, the 
reader will need to take some notation on trust and refer to later 
sections for the details.

We work on an interval $S_L=[-L/2,L/2]$ which we will think of as a 
circle of circumference $L$.  We let $P_{\pm}$ be the spectral projections 
of $-i\frac{\partial}{\partial x}$ regarded as a self adjoint operator on a 
dense domain in $L^2(S_L)$.  We let $\cF$ denote the free fermion Fock 
space over $L^2(S_L)$.  We choose the usual positive energy condition that 
the fermion fields are in a Fock representation of the algebra of the 
canonical anticommutation relations defined by $P_-$.  This means the 
fermion fields $\{\psi(f), \psi(g)^*\ | \ f,g\in L^2(S_L)\}$ 
satisfy
\eq
\left< \Omega, \psi(f)^* \psi(g)\Omega\right>_{\cF} =
\left< g, P_-f\right>_{L^2(S_L)}
\eqend
where $\Omega$ is the vacuum or cyclic vector in $\cF$.
We let $Q$ denote the Fermion charge operator on  $\cF$, and $R$ a 
unitary charge shift operator on $\cF$ satisfying $R^{-1}Q R = Q+I$
(the precise choice for $R$ will be explained later). 

We will construct regularised anyon field operators $\phi_\eps^\nu(x)$ 
where $\nu\in\R$ is a parameter determining the statistics, $x\in 
S_L$, and $\eps>0$ is a regularization parameter.  For positive $\eps$ 
the operator $\phi_\eps^\nu(x)$ is proportional to a unitary operator 
on $\cF$ which represents a certain $\U(1)$ valued loop on $S_L$.  
These operators are not periodic but obey (the parameter $\nu_0$ will 
be explained below),
$$
\phi_\eps^\nu(x+L) = e^{-i\pi\nu\nu_0 Q}
\phi_\eps^\nu(x)e^{-i\pi\nu\nu_0 Q}, 
$$
and in the limit as $\eps\downarrow 0$ they converge to operator valued 
distributions $\phi^\nu(x)$ satisfying
\eq
\label{exc}
\phi^\nu(x)\phi^{\nu'}(y) =e^{-i\pi\nu\nu'\sgn(x-y)}
\phi^{\nu'}(y)\phi^\nu(x).  
\eqend
In particular for $p \in \Lambda^*=\{\tPiL n\vert n\in \Z\}$, 
the formula
\eq
\label{phinup}
\hat\phi^\nu(p)=\lim_{\eps\downarrow 0}\int_{-L/2}^{L/2}dx\, e^{ipx} 
e^{i\pi\nu\nu_0 Q x /L} \phi_\eps^\nu(x)e^{i\pi\nu\nu_0 Q x/L} \eqend 
is a well-defined operator on $\cF$ (Proposition 1).  Note that we 
have to insert factors to compensate for the non-periodicity of 
$\phi_\eps^\nu(x)$ before Fourier transformation.  We also find that 
the statistics parameters $\nu,\nu'$ for which Eq.\ \Ref{exc} holds 
cannot be arbitrary but have to be integer multiples of some fixed 
(arbitrary) number $\nu_0>0$.  (If one is only interested in a single 
species of anyons one can chose $\nu_0=|\nu|$.)

A main focus is on the correlation functions of the anyon fields.
These are distributions defined by taking the limit as $\eps_j\downarrow 
0$ of 
\eq
\label{corr}
C_{\eps_1,\ldots,\eps_N}^{\nu_1,\ldots,\nu_N}(\nu_0,w_1,w_2|y_1,\ldots,y_N):=
\left<\Omega, R^{w_1} \phi_{\eps_1}^{\nu_1}(x_1)\cdots
\phi_{\eps_N}^{\nu_N}(x_N) R^{w_2}\Omega\right> 
\eqend
for $x_j\in S_L$, $\nu_j/\nu_0 \in \Z$ (for fixed $\nu_0$). 
Using general results for implementors of $\U(1)$ loops we obtain 
\eqa
\label{corr1}
C_{\eps_1,\ldots,\eps_N}^{\nu_1,\ldots,\nu_N}(\nu_0,w_1,w_2|y_1,\ldots,y_N)=
\delta_{w_1+w_2 +(\nu_1+\ldots +\nu_N)/\nu_0,0}\nonu \times 
e^{i\pi(w_1-w_2)\nu_0(\nu_1x_1+\cdots \nu_N x_N)/L}
\prod_{j=1}^N \prod_{k=j+1}^N  
b(x_j-x_k; \eps_j+\eps_k)^{\nu_j\nu_k}
\eqaend
with
\eq 
\label{b}
b(x,\eps):= \left( e^{-i \PiL x} - e^{-\tPiL \eps} e^{i\PiL x} \right) 
= -2ie^{-\pi\eps/L}\sin\PiL(x+i\eps).  \eqend The reason for studying 
these correlation functions is the connection with the 
Calogero-Sutherland (CS) Hamiltonian \cite{Su}.  This is defined on 
the set of functions $ f\in C^2(S_L^N;\C) $ which are zero on 
$\{(x_1,\ldots,x_N)\in S_L^N | \; \mbox{$x_j=x_k$ for some $k\neq j$ 
and/or $x_j=\pm L/2$ } \}$, \eq
\label{Sutherland}
H_{N,\beta}= - \sum_{j=1}^N\frac{\partial^2}{\partial x_j^2} 
+\sum_{\stackrel{j,k=1}{j\neq k}}^N 
\frac{(\PiL)^2\beta(\beta-1)}{\sin^2\frac{\pi}{L}(x_j-x_k)} , \eqend 
and which extends to a self-adjoint operator on 
$L^2(S_L^N)$.\footnote{Since Eq.\ \Ref{Sutherland} obviously is a 
positive symmetric operator, this follows e.g.\ from Theorem X.23 in 
Ref.\ \cite{RS2} (the Friedrich's extension).  Our approach will lead 
to a particular self-adjoint extension which is related to the 
standard one \cite{Su} in a simple manner.}

We will prove that the eigenfunctions and spectrum of this 
Hamiltonian 
can be obtained from anyon correlation functions,
namely as finite linear combinations of functions
\eq
\label{fnu}
f_{\nu,N}(\vn|\vx) := \lim_{\eps\downarrow 0}\left<\Omega, 
\hat\phi^\nu(\tPiL n_N)^*\cdots \hat\phi^{\nu}(\tPiL 
n_1)^*\phi_\eps^{\nu}(x_1)\cdots \phi_\eps^{\nu}(x_N) \Omega\right> 
\eqend where $n_j\in\N_0$ (Theorem 3).  We will obtain these results 
by constructing a self-adjoint operator $\cH^{\nu,3}$ which can be 
regarded as a `second quantization' of the CS Hamiltonian: it 
obeys the relations
$$
\cH^{\nu,3} \phi^\nu_\eps(x_1)\cdots \phi^\nu_\eps(x_N) \Omega \simeq
H_{N,\nu^2}\phi^\nu_\eps(x_1)\cdots \phi^\nu_\eps(x_N) 
\Omega
$$
 where `$\simeq$' mean `equal in the limit $\eps\downarrow 0$' (see 
 Theorem 2 for details).  We obtain $\cH^{\nu,3}$ by arguing by 
 analogy with the well known $W$-algebra associated with fermions.  
 Using analogous formulae we construct the first few generators 
 $\cH^{\nu,s}$,  $s=1,2,3$, of an anyon $W$-algebra.  Understanding the 
 complete anyon $W$-algebra is a problem we leave for a further 
 investigation.

This main result implies explicit formulas for the eigenvalues and a 
simple algorithm to construct eigenvectors $\Psi_{\nu,N}(\vn)  $ of 
$H_{N,\nu^2}$ as finite linear combinations of vectors
\eq
\label{eta}
\eta_{\nu,N}(\vn) 
=\hat\phi^\nu(\tPiL n_1)\cdots \hat\phi^\nu(\tPiL n_N) \Omega,
\quad \vn=(n_1,\ldots,n_N)\in\N_0^N . 
\eqend
These vectors Eq.\ \Ref{eta} can be naturally interpreted as $N$-anyon 
states with anyon momenta $p_j=\tPiL n_j$. 
Using these relations we can compute 
$$\left<\Psi_{\nu,N}(\vn) , 
\cH^{\nu,3} \phi^\nu_\eps(x_1)\cdots \phi^\nu_\eps(x_N) \Omega\right>
$$
in two different ways, and in the limit $\eps\downarrow 
0$ we obtain functions (of the variables $(x_1,\ldots, x_N)$) in $L^2(S^N_L)$ 
which are the promised eigenfunctions of $H_{N,\nu^2}$ (Theorem 3).

In the last subsection we observe that we recover the known spectrum 
of the CS Hamiltonian.  Comparing with the known solutions of 
the CS model \cite{Fo2}, we can establish the relationship 
between the eigenfunctions of Theorem 3 and the Jack polynomials.

\section{Preliminaries}

The subsequent discussion relies
on some standard material which is summarized in this section.  We will follow 
essentially the treatment in \cite{CH},\cite{PS},\cite{CR}.

\subsection{Notation} 
We denote by $\N$ and $\N_0$ the positive and non-negative integers, 
respectively. 
Let 
\eq
\Lambda^*=\{p=\frac{2\pi}{L}n\quad \vert\quad n\in \Z\}
\eqend
and 
\eq
\Lambda^*_0=\{k=\frac{2\pi}{L}(n+\frac{1}{2})\quad \vert\quad n\in \Z\}.
\eqend
Our underlying Hilbert space for the fermions we take to be
$L^2(S_L)\cong \ell^2(\Lambda^*_0)$.
These are identified via the Fourier transform defined by
\eq
\hat f(k) = \frac{1}{\sqrt{2\pi}}\int_{-L/2}^{L/2} dx f(x) e^{-ikx}
\eqend
for $k\in \Lambda^*_0$. 
An orthogonal basis of $L^2(S_L)$
is provided by the functions 
\eq
\label{ek}
e_k(x)= \frac{1}{\sqrt{2\pi}}e^{ikx}, \quad k\in \Lambda^*_0.
\eqend
and then we have
$$
f=\frac{2\pi}{L} \sum_k\hat f(k)e_k.
$$
The spectral projection $P_-$ corresponding to the negative 
eigenvalues of $\frac{1}{i}\frac{\partial}{\partial x}$ is defined as 
$\widehat{(P_-f)}(k)=\hat f(k)$ for $k<0$ and $=0$ otherwise.  We also 
use $P_+=I-P_-$.

\subsection{Quasi-free representations of the CAR algebra}
Let
$\{a(f), a(g)^*\ \vert \ f,g \in L^2(S_L)\}
$
be the usual generators of the fermion field algebra over $L^2(S_L)$, 
satisfying the canonical anticommutation relations (CAR) 
\eq
a(f)a(g)+a(g)a(f)=0,\quad a(f)a(g)^*+a(g)^*a(f)=
\left< f, g\right>_{L^2(S_L)} I.
\eqend
In the representation $\pi_{P_-}$ of this algebra determined by the 
projection $P_-$ we write $\psi(f)=\pi_{P_-}(a(f))$. If $\Omega$ denotes the 
cyclic (or vacuum) vector in the Fock space $\cF$ on which $\pi_{P_-}$ 
acts then this representation is specified by 
the following conditions, 
\eq
\psi(P_+f)\Omega=0=\psi^*(P_-f)\Omega .
\eqend
We also use the notation
\eq
\label{psik}
\hat\psi^{(*)}(k)=\psi^{(*)}(e_k),\quad k\in\Lambda_0^*.
\eqend

\subsection{Wedge representation of the loop group}
Each unitary operator $U$ on $L^2(S_L)$, with $P_\pm UP_\mp$ 
Hilbert-Schmidt, defines an `implementer' $\Gamma(U)$, on the Fock 
space $\cF$ satisfying \eq \Gamma(U)\psi(f)\Gamma(U)^{-1}= \psi(Uf).  
\label{3.1} \eqend Of particular interest is the representation of the 
smooth loop group ${\cG}=C^\infty(S_L;U(1))$ of $U(1)$ by implementors 
of the unitaries $U(\varphi)$ acting on $L^2(S_L)$.  These are defined 
for $\varphi\in \cG$ by \eq U(\varphi)f= \varphi f, \quad f\in 
L^2(S_L).  \eqend Then $\Gamma$ gives a projective representation of 
$\cG$ on $\cF$.  Writing $\Gamma(\varphi)$ for $\Gamma(U(\varphi))$ we 
may choose
 \eq \Gamma(\varphi)^*=\Gamma(\varphi^*) \eqend and we have 
 \eq
\Gamma(\varphi)\Gamma(\varphi')= \sigma(\varphi, \varphi')
\Gamma(\varphi\varphi')  \label{3.2}
\eqend
where $\sigma(\varphi, \varphi')$ is some $U(1)$ valued group two-cocycle
on $\cG$. We will determine this cocycle next.

The choice of phase of $\Gamma(\varphi)$ is important
for giving an exact formula for $\sigma$.  For those 
$\varphi=e^{i\alpha}$, with $\alpha\in Lie\cG:=C^\infty(S_L;\R)$, the map $r\to 
\Gamma(e^{ir\alpha})$ is required to be a one parameter group such that the 
generator $d\Gamma(\alpha)$ of this group satisfies $\left<
\Omega,d\Gamma(\alpha)\Omega\right> = 0.$ Then we have 
\eq
\label{3.3b}
[d\Gamma(\alpha), \psi(g)^*]= \psi(\alpha g)^*  \label{3.3}
\eqend
and a standard calculation \cite{CH},\cite{PS},\cite{CR}
gives
\eq
[ d\Gamma(\alpha_1),d\Gamma(\alpha_2) ]= i s(\alpha_1,\alpha_2) I
\label{3.4}
\eqend
with 
the Lie algebra two-cocycle
\eq
 s(\alpha_1,\alpha_2)=\frac{1}{4\pi}\int_{-L/2}^{L/2} 
dx (\frac{d\alpha_1(x)}{dx}\alpha_2(x) -\alpha_1(x) 
\frac{d\alpha_2(x)}{dx}). \label{3.5}
\eqend
Hence the
\eq
\label{dGamma}
\Gamma(e^{i\alpha}) = e^{id\Gamma(\alpha)}
\eqend
are Weyl operators satisfying Eq.\ (3.2) with 
$
\sigma (e^{i\alpha_1}, e^{i\alpha_2})=
e^{- i s(\alpha_1,\alpha_2)/2 } .
$

We will also use $d\Gamma(\alpha)$ for complex valued $\alpha$.  These are 
naturally defined by linearity, 
\eq
\label{lin}
 d\Gamma(\alpha_1+i\alpha_2) = d\Gamma(\alpha_1)+ i d\Gamma(\alpha_2) \quad
\alpha_{1,2}\in C^\infty(S_L;\R). 
\eqend
Then 
\eq
d\Gamma(\alpha)^*=d\Gamma(\alpha^*)  \label{3.6}
\eqend 
(we use the same symbol $*$ for Hilbert space adjoints and complex 
conjugation), and Eqs.\ \Ref{3.4} and \Ref{3.5} extend to $C^\infty(S_L;\C)$ so 
that $s$ defines a complex bilinear form in an obvious way.

Here a technical remark is in order.  The operators $d\Gamma(\alpha)$, 
$\alpha\in C^\infty(S_L;\C)$ are all unbounded.  However, there is a 
common, dense, domain $\cD$ which is left invariant by all operators 
$\Gamma(\varphi)$, $\varphi\in\cG$ (this is discussed in more detail 
in Appendix B).  Thus Eqs.\ \Ref{3.3b}, \Ref{3.4}, \Ref{lin} and 
similar equations below are all well-defined on $\cD$.  We also note 
that all vectors in $\cD$ are analytic for all the operators 
$d\Gamma(\alpha)$, $\alpha\in C^\infty(S_L;\C)$ (see e.g.\ \cite{CR}).

It is convenient to decompose loops into their positive, negative and zero 
Fourier components, 
\eqa
\alpha(x)=\alpha^+(x)+\alpha^-(x)+\bar\alpha;
\ \alpha^\pm(x) = 
\frac{1}{L}\sum_{\pm p>0} \hat\alpha(p) e^{ipx}, \quad 
\bar\alpha = \frac{1}{L} \hat\alpha(0)  \label{3.7}
\eqaend 
where
\eq
\hat\alpha(p)= \int_{-L/2}^{L/2} dx \, \alpha(x) e^{-ipx}\quad 
p\in\Lambda^*. 
\eqend
Then
\eq
d\Gamma(\alpha)=d\Gamma(\alpha^+)+d\Gamma(\alpha^-) + \bar\alpha Q
\eqend
with $Q=d\Gamma(I)$. Note that
\eq
d\Gamma(\alpha^-)\Omega=d\Gamma(\alpha^+)^*\Omega=Q\Omega=0 
\label{3.8}
\eqend
(highest weight condition) implying 
\eq
\left< \Omega, d\Gamma(\alpha_1)d\Gamma(\alpha_2)\Omega\right>
=
\left< \Omega, d\Gamma(\alpha_1^-)d\Gamma(\alpha_2^+) \Omega\right>
= i s(\alpha_1^-,\alpha_2^+).
\eqend
We also have $i s(\alpha^-,\alpha^+)\geq 0$
which can be also easily be seen from the explicit formula for $s$, 
\eq
 i s(\alpha_1,\alpha_2)= \sum_{p\in\Lambda^*} 
\frac{p}{2\pi L} \hat\alpha_1(-p)\hat\alpha_2(p) . 
\eqend 
Standard arguments now give us (for $\alpha$ real valued) 
\eq
\label{delta}
\left<\Omega,\Gamma(e^{i\alpha})\Omega\right> = e^{-i 
s(\alpha^-,\alpha^+) } . 
\eqend
We also need $R=\Gamma(\phi_1)$ which implements the operator 
$U(\phi_1)$ where $\phi_1(x)=e^{2\pi ix/L}$ (for an explicit construction 
of $\Gamma(\phi_1)$ see e.g.\ \cite{Ruij}).  The phase of this unitary 
operator will be fixed latter.  Notice that 
\eq
R^{-1} d\Gamma(\alpha) R = d\Gamma(\alpha)+\bar\alpha I .   
\label{3.10}
\eqend 
(this will be explained in more detail in Appendix A). 

General loops in $\cG$ are of the form $\varphi=e^{if}$ with 
\eq
f(x)= w 
\frac{2\pi}{L} x + \alpha(x)  \label{3.11}
\eqend
with periodic $\alpha$ and integer 
$w=[ f( L/2 )-f( -L/2 ) ]/2\pi$ ($w$ is the winding number of 
$\varphi$).  We then define 
\eq
\Gamma( e^{if} ) : = e^{i \bar\alpha Q/2} R^{w} e^{i \bar\alpha Q/2} 
\Gamma(e^{i(\alpha^+ +\alpha^-)}) . \label{3.12}
\eqend
This fixes the phase for all implementors. With that we get 
\eq
\label{sgma}
\sigma( e^{if_1}, e^{if_2} ) = e^{-  i S (f_1,f_2)/2} \label{3.13}
\eqend
where we introduced 
\eq
\label{S1}
 S(f_1,f_2)= s(\alpha_1,\alpha_2) + ( 
w_{f_1}\bar\alpha_2 - \bar\alpha_1 w_{f_2} ).  
\eqend 
It is worth 
noting that one can write
\eqa
S(f_1,f_2) = f_1(\frac{L}{2})f_2(-\frac{L}{2}) 
- f_1(-\frac{L}{2})f_2(\frac{L}{2}) 
+ \frac{1}{4\pi} \int_{-L/2}^{L/2} dx 
(\frac{d f_1(x)}{dx}f_2(x) -f_1(x) \frac{d f_2(x)}{dx}) 
\eqaend 
which (up to trivial, but nevertheless important, rescaling of 
variables) is identical to the antisymmetric two cocycle introduced by 
Segal \cite{Seg}.  Notice that our choice of phase for the 
implementors implies that \eq
\label{delt}
\left<\Omega,\Gamma(e^{i f})\Omega\right> = 0\quad \mbox{ if $w\neq 0$} . 
\eqend

We will need the following relation
\eq
\Gamma(e^{if_1})\Gamma(e^{if_2}) \cdots \Gamma(e^{if_N})  
= (\prod_{j<k}
e^{- iS(f_j,f_k)/2 } )  \Gamma(e^{if_1} e^{if_2} \cdots e^{if_N}) 
\label{3.15}
\eqend 
which follows by induction (here and in the following
$\prod_{j<k}$ is short for $\prod_{j=1}^N \prod_{k=j+1}^N$). 

We introduce normal ordering $\;\xxa \cdots \xxe\;$ as follows. 
For implementors of loops of winding number zero it is defined as 
\eq
\label{xx1}
\xxa\Gamma(e^{i\alpha})\xxe\;  : = e^{ i S(\alpha^-,\alpha^+)/2} 
\Gamma(e^{i\alpha})
\eqend
with the numerical factor chosen such that 
$\langle\Omega,\xxa\Gamma(e^{i\alpha})\xxe\Omega\rangle=1
$ [cf.\ Eq.\ \Ref{delta}] .  We extend this to implementors of 
general loops,
\eq
\label{xx11}
\xxa\Gamma(e^{i f})\xxe\;  : = e^{i\bar \alpha Q/2} R^w e^{i\bar \alpha Q/2} 
\xxa\Gamma(e^{i(\alpha^+ + \alpha^-) })\xxe
\eqend
and to products of implementors, 
\eq
\label{xx2}
\xxa \Gamma(e^{if_1}) \Gamma(e^{if_2}) \cdots \Gamma(e^{if_N})\xxe \; := \; 
\xxa\Gamma(e^{if_1}e^{if_2}\cdots e^{if_N}) \xxe .   
\eqend
A straightforward computation then implies the
following relations
\eq
\label{noo}
\xxa \Gamma(e^{if_1})\xxe \xxa \Gamma(e^{if_1})\xxe \; = 
e^{-i\tilde S(f_1,f_2)/2 }
\xxa \Gamma(e^{if_1})\Gamma(e^{if_1})\xxe
\eqend
with 
\eq
\label{tildeS}
\tilde S(f_1,f_2) = w_1\bar\alpha_2 - \bar\alpha_1 w_2 
+2S(\alpha_1^-,\alpha_2^+) = -\tilde S(f_2,f_1)^* 
\eqend
which will be useful in the following. Finally,  
\eq
\label{normal ordering}
\xxa d\Gamma(\alpha_1)\cdots d\Gamma(\alpha_m) \Gamma(e^{if}) \xxe\; := 
(-i)^m\left.
\frac{\partial^m}{\partial a_1\cdots \partial a_m}
\xxa e^{ia_1 d\Gamma(\alpha_1)}\cdots e^{ia_m d\Gamma(\alpha_m)} 
\Gamma(e^{if}) 
 \xxe  \right|_{a_j=0}  
\eqend
and 
\eq
\xxa AB\xxe\;=\;\xxa BA\xxe\;=\;\xxa (\xxa A\xxe) B\xxe\;=\; 
\xxa (\xxa A\xxe) (\xxa B \xxe) \xxe 
\eqend
extends the definition of normal ordering to arbitrary products of 
operators $d\Gamma(\alpha_j)$ and $\Gamma(e^{if_k})$. We note that
by Stone's theorem \cite{RS} the 
differentiations here are well-defined in the strong sense on the dense
domain $\cD$ defined in Appendix B. 

It is convenient to introduce the operators
\eq
\label{bosons}
\b(p) :=d\Gamma(\epsilon_p), \quad \epsilon_p(x)=e^{-ipx},\quad 
p\in\Lambda^* \eqend which allow us to write \eq 
d\Gamma(\alpha)=\sum_{p\in\Lambda}\hat\alpha(p)\b(-p) .  \eqend The 
$\b(p)$ have a natural interpretation as boson field operators and 
will be further discussed in Appendix A. The subspace $\cD_b$ (finite 
boson vectors) of $\cF$ spanned by vectors of the form \eq
\label{etab}
\eta_b=\b(-q_1)\cdots \b(-q_n)R^\ell\Omega,\quad q_j>0, 
n\in\N_0,\ell\in\Z 
\eqend
will be important for us. Note that $\cD_b$ is dense in $\cF$ 
(see e.g.\ \cite{CR}). 

\subsection*{Appendix A. Relation to quantum field theory}

In this section we make contact with notation from the more
algebraic approach to the results summarized above
\cite{Kac}, \cite{KR}. 
This notation is close to that
commonly used in the physics literature.
First  the representation $\pi_{P_-}$ 
of the CAR algebra can be described in terms of the operators 
$\hat\psi^{(*)}(k)$ Eq.\ \Ref{psik} which satisfy the following relations 
\eq
\hat\psi(k)\hat\psi(k')^* + \hat\psi(k')^*\hat\psi(k) 
= \frac{L}{2\pi}\delta_{k,k'} I
\eqend
and					
\eq
\label{hw1}
\hat\psi(k)\Omega=0=\hat\psi^*(-k)\Omega, \quad k>0.
\eqend

We chose the physics notation for 
 the operators $\b(p)$ defined in Eq.\ \Ref{bosons}. These operators
satisfy some additional relations easily proved from their definition. 
For example, Eqs.\ \Ref{3.3}--\Ref{3.6} imply
$[\b(p), \hat\psi(k)^*]= \hat\psi(k-p)^*, $
\eq
\quad [\b(p),\b(q)] = p\frac{L}{2\pi}\delta_{-p,q}I, \quad p,q\in 
\Lambda^*, \label{3.19}
\eqend
and $\b(-p)=\b(p)^*$. Moreover, 
\eq
\label{bOm}
\b(p)\Omega=0\quad p\geq 0  
\eqend
follows from Eq.\ \Ref{3.8}. 
If we define the usual Wick ordering for free fermions by 
\eq
:\hat\psi(k)^*\hat\psi(k^\prime):=
\left\{\begin{array}{ll}  -\hat\psi(k^\prime)\hat\psi(k)^*\quad & 
\mbox{ if $k^\prime=k<0$} \\
\quad \hat\psi(k)^*\hat\psi(k^\prime)\quad & \mbox{ 
otherwise,}\end{array}\right.  \eqend we can write \eq 
\b(p)=\frac{2\pi}{L}\sum_{k\in \Lambda^*_0} 
:\hat\psi^*(k-p)\hat\psi(k): \label{3.20}.  \eqend Since this formally 
is equivalent to $\b(p)=\int_{S_L}dx \, :\psi^*(x)\psi(x): e^{-ipx}$ 
the $\b(p)$ can be interpreted as the Fourier modes of the fermion 
currents which (formally) are defined as $\rho(x)= \, 
:\psi^*(x)\psi(x):$.  This motivates our notation for these operators.  
In particular $Q=\b(0)$ is the fermion charge operator.

It follows from Eq.\ \Ref{3.1} and the definition of $R$ that 
\eq
R\hat\psi(k)R^{-1} 
=\psi(k+ \frac{2\pi}{L}).
\eqend 
{}From (58) and \Ref{3.19} we deduce the important 
relation: 
\eq
R^{-1}\b(p)R = \b(p) +\delta_{p,0}I
\eqend 
equivalent to Eq.\ \Ref{3.10}. 
This implies $R^{-w}QR^w=Q+w I$ for arbitrary integers $w$. 
Notice that as $R^{w}\Omega$ is in the eigenspace of $Q$ with eigenvalue
$w$ we have $\left< \Omega, R^w \Omega\right>=\delta_{w,0}$. 
More generally, $(Q - w I)\Gamma(e^{if})\Omega=0$, which 
implies Eq.\ \Ref{delt}.

\subsection*{Appendix B. Domains for unbounded operators}

In this paper we are dealing with an algebra of
unbounded operators. For many of the subsequent calculations
to make mathematical sense it is essential
to understand how the domain on which they all act is obtained.
The technical results we need are all contained in \cite{CR,GL}.

As mentioned above, implementers $d\Gamma(\alpha)$ of loops $\alpha\in 
C^\infty(S_L;\C)$ are unbounded.  However they have a common invariant 
dense domain $\cD$ which we now describe. For vectors
\eq
\label{etaf}
\eta_f= \hat\psi^*(k_1)\cdots \hat\psi^*(k_n)\hat\psi(-\ell_1)\cdots 
\hat\psi(-\ell_m)\Omega, \quad k_i,\ell_j>0, n,m\in\N_0 
\eqend
we set
\eq
P_\lambda \eta_f :=\left\{\bma{cc} \eta_f& \mbox{ if $n+m\leq \lambda $} 
\\ 0& \mbox{ otherwise} 
\ema \right. ,\quad \lambda \in \N ,
\eqend
and this defines a family of projection operators on $\cF$ such that 
$s-\lim_{\lambda\to\infty}P_\lambda=I$; see e.g.\ \cite{CR}. 
Thus
\eq
{\cD_0} : = \{F\in{\cF}\vert P_\lambda F =F 
\mbox{ for some positive integer $\lambda$ }\}
\eqend 
is a dense subspace in $\cF$.  The space $\cD_0$ consists of analytic 
vectors for the operators $d\Gamma(\alpha)$, $\alpha\in C^\infty(S_L;\C)$.  
This follows from $$ d\Gamma(\alpha)P_\lambda = P_{\lambda+2} 
d\Gamma(\alpha)P_\lambda , \quad || d\Gamma(\alpha)P_\lambda ||\leq 
C_\alpha(\lambda+2) $$ where $||\cdots||$ is the operator norm and 
$C_\alpha$ a constant depending only on $\alpha$, see \cite{CR}.  It 
follows that $d\Gamma(\alpha)$, $\alpha\in C^\infty(S_L,\R)$, is 
essentially self-adjoint on $\cD_0$.

We extend $\cD_0$ to a space which is also invariant under all implementers 
of the loop group and define $\cD$ as the linear span of vectors 
$\Gamma(\varphi)F$, $F\in\cD_0$ and $\varphi\in\cG$.  We summarize the 
properties of the domain $\cD$: 
\vspace*{0.45cm} 

\noindent
{\em 
$(i)$ $\cD$ is a common, dense, invariant set of analytic vectors 
for all operators $d\Gamma(\alpha)$, $\alpha\in C^\infty(S_L,\C)$,
\\ \noindent
$(ii)$ $\cD$ is invariant under all operators $\Gamma(\varphi)$, 
$\varphi\in\cG$,
\\ \noindent
$(iii)$ $\cD$ contains $\cD_b$. 
}

\vspace*{0.45cm} 

\noindent
(The properties $(i)$ and $(ii)$ follow from the corresponding properties of 
$\cD_0$ and the following relation,
\eq
\label{inv}
\Gamma(e^{-if})d\Gamma(\alpha) \Gamma(e^{if})=d\Gamma(\alpha)+S(f,\alpha)I
\eqend
which is easily proved using Eqs.\ \Ref{3.2}, \Ref{dGamma}, 
\Ref{3.13} and 
\Ref{lin}. Property $(iii)$ follows trivially from the definitions.)

We finally justify a formula which we will need below.  We observe 
that {\em formally}, $\xxa\Gamma(e^{i f})\xxe$ Eq.\ \Ref{xx11} equals
$$
e^{i\bar \alpha Q/2} R^w e^{i\bar \alpha Q/2} 
e^{id\Gamma(\alpha^+)} e^{id\Gamma(\alpha^-)} . 
$$
(using $e^{i(A_+ +A_-)}=e^{i A_+}e^{i A_-}e^{[A_+,A_-]/2}$ for 
$A_\pm=d\Gamma(\alpha^\pm)$ and Eq.\ \Ref{3.4} {\em ff}, this would 
account for Eq.\ \Ref{xx1}).  This formula is problematic since the 
operators $e^{id\Gamma(\alpha)}$ are only defined for {\em real}-valued 
functions $\alpha$.  However, $$
\left<\Omega,  R^{-\ell}\b(k_m)\cdots\b(k_1) d\Gamma(\alpha^\pm)^n
\b(-k_1')\cdots \b(-k'_{m'})R^{\ell'}\Omega\right>
$$
is always zero for $n>\max(m,m')$ (this is easily proved by using 
Eq.\ \Ref{3.8} after applying repeatedly Eq.\ \Ref{3.4}). Thus
$$
e^{id\Gamma(\alpha^\pm)} := \sum_{n=0}^\infty 
\frac{i^n}{n!}d\Gamma(\alpha^\pm)^n
$$
can be defined as a sesquilinear form on $\cD_b$.  Since 
$e^{id\Gamma(\alpha^-)}R^\ell\Omega =R^\ell\Omega$ for all 
$\ell\in\Z$, it follows that
 \eq
\label{imp}
\Gamma(e^{if}) R^\ell\Omega = 
e^{i\bar\alpha (w/2 +\ell) } \sum_{n=0}^\infty
\frac{i^n}{n!}d\Gamma(\alpha^+)^n R^{w+\ell}\Omega
\eqend
where the r.h.s of this equation is well-defined as an element in the dual 
of $\cD_b$.

\section{Vertex Operators}

\subsection{Boson-fermion correspondence}
As a motivation and to introduce notation, we first recall how the 
boson-fermion correspondence can be derived from the results summarized in 
the last Section
\cite{PS,CH}. In Ref.\ \cite{Seg} a so-called `blip' function was 
introduced which  equals, up to the sign, 
$$
\frac{ e^{i(x-y)2\pi/L} -\lambda }{ 1- \lambda e^{i(x-y)2\pi/L}} , \quad 
0<\lambda<1 . 
$$
which is the exponential of a smoothed out step 
function.  Writing it as $e^{ i f_{y,\eps} }$ with $\lambda=e^{-
2\pi\eps / L }$ one gets 
\eq
\label{f}
f_{y,\eps}(x) = \frac{2\pi}{L}(x-y) +
\alpha^+_{y,\eps}(x) +\alpha^-_{y,\eps}(x) 
\eqend
with
\eq
\label{alpha}
\alpha^\pm_{y,\eps}(x) = \pm i \log(1- e^{2\pi (\pm i(x-y)-\eps)/L } )
=\mp i\sum_{n=1}^\infty \frac{1}{n} e^{\pm 2i\pi n(x-y)/L} e^{-2\pi\eps n/L}. 
\eqend
Note that the winding number of $f_{y,\eps}$ equals $1$. 
Since $f_{y,\eps}(x)$ for $\eps\downarrow 0$ converges to $ \pi 
\sgn(x-y)$ we will also use the following suggestive notation,
\eq
\label{sgn}
\sgn(x-y;\eps) := \frac{1}{\pi} f_{y,\eps}(x) . 
\eqend
Later we will also need the function 
$\delta_{y,\eps}(x)= \partial_x f_{y,\eps}(x)/2\pi$ i.e.\ 
\eq
\label{delta1}
\delta_{y,\eps}(x)= \frac{1}{L} + \delta^+_{y,\eps}(x) + \delta^-_{y,\eps}(x)
\eqend
with 
\eq
\label{deltapm}
\delta^\pm_{y,\eps}(x)  
=\frac{1}{L}\sum_{n>0} e^{\pm  2\pi i (x-y)n/L} e^{-2\pi\eps n /L }. 
\eqend

These functions have the following important properties which we summarize as 
\vspace*{0.45cm} 

{\bf \noindent  Lemma 1:} 
\eqa
\label{blips}
S(\alpha^-_{y,\eps},\alpha^+_{y',\eps'}) &=& 
\alpha^+_{y',\eps+\eps'}(y) \nonu
S(f_{y,\eps},f_{y',\eps'}) &=&  \pi \sgn(y-y'; \eps+\eps') \\
S(\delta^\mp_{y,\eps},\alpha^\pm_{y',\eps'}) &=& 
-\delta^\pm_{y',\eps+\eps'}(y)  \nonumber 
\eqaend
(The proof of these relations is a straightforward calculation which we skip.)

\vspace*{0.45cm} 

Then for $\eps>0$ and integer $\nu$ the operators 
$
\phi^{\nu}_\eps(y):= \; \xx \Gamma(e^{i \nu f_{y,\eps} }) \xx 
=\phi^{-\nu}_\eps(y)^*
$
are well-defined, and from Lemma 1 and Eqs.\  \Ref{3.2} and \Ref{sgma} 
we conclude
\eq
\label{exc1}
\phi^{\nu}_\eps(y)\phi^{\nu'}_{\eps'}(y') = e^{-i \pi  \nu \nu'
\sgn(y-y';\eps+\eps')   } 
\phi^{\nu'}_{\eps'}(y')\phi^{\nu}_\eps(y). 
\eqend
For odd integers $\nu,\nu'$ and in the limit 
$\eps,\eps'\downarrow 0$ these formally become anticommutator 
relations. This suggests that the $\phi^{\pm 1}_\eps(y)$ in this limit 
are fermion operators. Indeed one can prove 
\eq
\label{bosonfermion}
\hat\psi^*(k)=\lim_{\eps\downarrow 0}\frac{1}{ \sqrt{2\pi L} }
\int_{-L/2}^{L/2} dy \, \phi^{1}_{\eps}(y) e^{iky}, \quad k\in \Lambda^*_0
\eqend
in the sense of strong convergence on a dense domain (see e.g.\ 
\cite{CH,PS}). 
This is the central result of the boson-fermion correspondence.  We note 
that this relation also fixes the phase of the unitary operator $R$.

\subsection{Construction of anyons}
To construct anyons we have to extend the relations Eq.\ \Ref{exc1} to 
non-integer $\nu,\nu'$.  However, the functions $e^{i\nu f_{y,\eps}(x)}$ 
are not periodic and thus $\Gamma(e^{i\nu f_{y,\eps}})$ does not exist.  To 
circumvent this problem, we note that $S(f_1,f_2)$ Eq.\ \Ref{S1} is 
invariant under changes $\bar \alpha_i\to \bar \alpha_i \lambda$ and 
$w_i\to w_i/\lambda$ with an arbitrary scaling parameter $\lambda$.  We use 
this to construct a function $\tilde f_{y,\eps}(x)$ which has the following 
properties,
$$
\bma{ll} (i)& \quad e^{i \nu \tilde f_{y,\eps}(x)}
\quad\mbox{ is periodic for all $\nu$,}\\
(ii)& \quad S(\tilde f_{y,\eps},\tilde f_{y,\eps})= 
S(f_{y,\eps},f_{y,\eps}).\nonumber
\ema
$$
Since the functions $\nu \tilde f_{y,\eps}(x)$ have winding numbers 
different from zero, the first requirement can only be fulfilled  
for $\nu$ values which are an integer multiple of some fixed number 
$\nu_0>0$.  Then
\eq
\label{tildef}
\tilde f_{y,\eps}(x) = 
\frac{2\pi}{L\nu_0}x -\frac{2\pi\nu_0}{L}y 
+ \alpha^+_{y,\eps}(x) +\alpha^-_{y,\eps}(x) 
\eqend
has the desired properties. Thus the 
operators
\eq
\label{anyon}
\phi_{\eps}^{\nu}(y): =\; \xxa\Gamma(e^{ i\mu \nu_0 \tilde 
f_{y,\eps}})\xxe\; =\phi_{\eps}^{-\nu}(y)^*, \quad \nu:=\nu_0\mu, 
\quad \mu\in\Z
\eqend 
are well-defined for $\eps>0$, and they obey the exchange relations Eq.\ 
\Ref{exc1} but now for all $\nu,\nu'$ which are integer multiples of $\nu_0$.
Thus the theory of loop groups provides a simple and rigorous construction 
of regularised 
anyon field operators $\phi_{\eps}^\nu(x)$.\\

{\em Remark:}
To be precise, one should denote the anyon operators defined in Eq.\ 
\Ref{anyon} as 
$\phi_{\eps}^{\nu_0,\mu}(y)$. 
Then Eq.\ \Ref{exc1} would read
$$
\phi^{\nu_0,\mu}_\eps(y)\phi^{\nu_0,\mu'}_{\eps'}(y') = e^{-i 
\pi\nu_0^2 \mu \mu'  
\sgn(y-y';\eps+\eps')   } 
\phi^{\nu_0,\mu'}_{\eps'}(y')\phi^{\nu_0,\mu}_\eps(y)\quad 
\mu,\mu'\in \Z.
$$
Making the $\nu_0$-dependence manifest would allow us to obtain 
slightly more general results. However, it would also lead to a 
proliferation of indices which is a price we are not willing to pay.\\ 

We note that this definition and Eq.\ \Ref{xx11} imply that the anyon 
fields are not periodic but the operators
\eq
\label{phinup2}
\check{\phi}_\eps^\nu(y):=
e^{i\pi\nu\nu_0 Q y /L} \phi_\eps^\nu(y)e^{i\pi\nu\nu_0 Q y/L} = 
R^{\nu/\nu_0} 
\xxa e^{i\nu d\Gamma (\alpha^+_{y,\eps}+\alpha^-_{y,\eps}) }\xxe 
\eqend
are.  This suggests that the Fourier modes $\hat\phi^\nu(p)$ of the anyons 
fields as defined in Eq.\ \Ref{phinup} are well-defined operators. 
In fact: 

\vspace*{0.45cm} 
{\bf \noindent Proposition 1:} {\em The $\hat\phi^\nu(p)$ defined in Eq.\
\Ref{phinup} are operators with $\cD_b$ as common, dense, invariant 
domain. Especially, 
\eq
\label{spec}
\hat\phi^\nu(0) R^\ell\Omega = R^{\ell +\nu/\nu_0}\Omega \quad \forall 
\ell\in\Z . 
\eqend
}
\vspace*{0.45cm} 

The proof of this is given in Appendix C. It implies that all vectors 
$\eta_{\nu,N}(\vn)$ Eq.\ \Ref{eta} are in $\cD_b$. This is important 
due to the following result also proven in Appendix C:

\vspace*{0.45cm} 
{\bf \noindent  Proposition 2:} {\em For $\eta\in\cD_b$, 
\eq
\label{Feta}
F^\nu_\eta(x_1,\ldots,x_N) : = \lim_{\eps\downarrow 0}\left<\eta, 
\phi_\eps^{\nu}(x_1)\cdots \phi_\eps^{\nu}(x_N) 
\Omega\right>
\eqend
exists and has the form
\eq
\label{Feta1}
F^\nu_\eta(x_1,\ldots,x_N) = e^{-i\pi\nu^2(x_1+\ldots +x_N)N/L 
}\Delta_{N}^{\nu^2}(x_1,\ldots,x_N)
\cP_{\eta}(\nu|e^{-2\pi i x_1/L},\ldots, e^{-2\pi i x_N/L})
\eqend
where
\eq
\label{Delta}
\Delta_{N}^{\nu^2}(\vx ) := \; \lim_{\eps\downarrow 0}  
\left( \prod_{j=1}^N\prod_{k=j+1}^N 
b(x_j-x_k;\eps) \right)^{\nu^2}
\eqend
with $b$ given in Eq.\ \Ref{b} 
and $\cP_\eta(\nu|\vz)$ a symmetric 
polynomial.\footnote{i.e.\ a polynomial which is invariant under 
permutations of the arguments; see  \cite{McD} . } 
Especially, $F^\nu_\eta(\vx) \in L^2(S_L^N)$.}
\vspace*{0.45cm}

Proposition 2 follows from the following explicit formula derived in 
Appendix C: for $\eta_b$ Eq.\ \Ref{etab}, 
\eq
\label{Fetab}
F^\nu_{\eta_b}(x_1,\cdots,x_N) = \delta_{\ell,N\nu/\nu_0}
e^{-i\pi\nu^2(x_1+\ldots 
+x_N)N/L } 
\prod_{j=1}^n\left( \sum_{k=1}^N \nu e^{-iq_j x_k  }   \right)
\Delta^{\nu^2}_{N}(x_1,\ldots,x_N) . 
\eqend
We note that $\Delta^{\nu^2}_{N}(x_1,\ldots,x_N)$ equals, up to a 
constant, to the well-known ground state wave function of the 
Sutherland model (see e.g.\ \cite{Su}). This will be further explored in Section 
\ref{suther}. 

Using Eqs.\ \Ref{eta} and \Ref{phinup} we now obtain
\eqa
\label{etanuN}
\eta_{\nu,N}(\vn) =
\lim_{\eps_1,\ldots,\eps_N\downarrow 0}
\int_{-L/2}^{L/2} dx_1 e^{ip_1x_1}
\cdots \int_{-L/2}^{L/2} dx_N e^{ip_N x_N}
\check\phi^{\nu}_{\eps_1}(x_1)\cdots \check
\phi^{\nu}_{\eps_N}(x_N)\Omega
\nonu = 
\lim_{\eps_1,\ldots,\eps_N\downarrow 0}
\int_{-L/2}^{L/2} dx_1 e^{iP_{1} x_1}
\cdots \int_{-L/2}^{L/2} dx_N e^{iP_{N} x_N}
\phi^{\nu}_{\eps_N}(x_1)\cdots \phi^{\nu}_{\eps_N}(x_N)\Omega
\eqaend
with $p_j=\frac{2\pi}{L}n_j$ and 
\eq
\label{Pj}
P_j = P_{j,\nu,N}(\vn) 
= \frac{2\pi}{L}\left(n_j +  \nu^2( N - j+ \half)  
\right)\quad j=1,2,\ldots N .
\eqend
(To derive this formula we used repeatedly $e^{icQ}R^w \Omega = 
e^{icw}R^w \Omega$ for $c\in\R$ and $w\in\Z$.) 
These $P_j$ can be interpreted as anyon momenta, and they will play 
an important role in Section \ref{suther}.  It is 
interesting to note how the momentum shifts $\propto \nu^2$ appear in our 
formalism: they are due to the factors $e^{-\pi\nu\nu_0 Qx/L}$ in Eq.\ 
\Ref{phinup} which are necessary to make the anyon operators periodic.

We finally formulate a {\em highest weight condition} for the Fourier 
modes of the anyon field operators which is analogous to Eq.\ \Ref{hw1} and 
will also play an important role in Section \ref{suther}. 

\vspace*{0.45cm}
{\bf \noindent  Proposition 3:} {\em The vector $\eta_{\nu,N}(\vn)$ Eq.\
\Ref{eta} is non-zero only if the following conditions are fulfilled,
\eqa
 n_1+n_2+\ldots + n_N  &\geq& 0 \label{conda} \\
n_{\ell}+\sum_{j=\ell+1}^N 2^{j-1-\ell} n_{j} &\geq& 0\quad \mbox{ for 
$\ell=1,2,\ldots N $} \label{cond} . 
\eqaend
}
\vspace*{0.45cm}

Again we defer the proof to Appendix C. 

\subsection{Anyon correlation functions} 
The results of the last two subsections enable us to 
complete one of our main aims namely
to compute all anyon correlations functions. First  
eqs. \Ref{xx1}, \Ref{xx2}, and \Ref{blips} imply 
\eq
\label{no}
\phi_{\eps_1}^{\nu_1}(x_1)\cdots
\phi_{\eps_N}^{\nu_N}(x_N) = 
\cJ^{\nu_1,\cdots,\nu_N}_{\eps_1,\cdots,\eps_N}(x_1,\ldots,x_N) 
\xxa \phi_{\eps_1}^{\nu_1}(x_1) \cdots
\phi_{\eps_N}^{\nu_N}(x_N)\xxe 
\eqend
where 
\eq
\label{J}
\cJ^{\nu_1,\cdots,\nu_N}_{\eps_1,\cdots,\eps_N}(x_1,\ldots,x_N) 
=
\prod_{j<k} b(x_j-x_k; \eps_j+\eps_k)^{\nu_j\nu_k}
\eqend
and the function $b(r,\eps)$ is defined in Eq.\ \Ref{b}.
Note that our definition of normal ordering implies
\eq
\label{vev}
\left<\Omega, R^{w_1} \xxa \phi_{\eps_1}^{\nu_1}(x_1) \cdots
\phi_{\eps_N}^{\nu_N}(x_N)\xxe R^{w_2}\Omega\right> = 
\delta_{w_1+w_2+(\nu_1+\ldots+\nu_N)/\nu_0,0}
e^{ i\pi(w_1-w_2)\nu_0(\nu_1x_1+\ldots+\nu_N x_N)/L} . 
\eqend

Now using equations \Ref{delta}, \Ref{delt} we obtain Eqs.\ 
\Ref{corr}--\Ref{corr1}.
Our main interest is in the functions \Ref{fnu} which 
can be written as
\eqa
f_{\nu,N}(\vn|\vx) = F^\nu_{\eta_{\nu,N}(\vn)} (\vx) . 
\eqaend
By a simple computation, 
\nonueqa
f_{\nu,N}(\vn|\vx) = 
\lim_{\eps\downarrow 0}
\lim_{\eps_1,\ldots,\eps_N\downarrow 0}\int_{-L/2}^{L/2} d y_1 e^{-iP_1y_1}
\cdots \int_{-L/2}^{L/2} dy_N e^{-iP_N y_N}\nonu\times
\left< \Omega, \phi^{-\nu}_{\eps_N}(y_N)\cdots 
\phi^{-\nu}_{\eps_1}(y_1)\phi^{\nu}_\eps(x_1)\cdots 
\phi^{\nu}_\eps(x_N)\Omega \right>
\nonu = 
e^{-i\pi\nu^2(x_1+\ldots +x_N)N/L 
}\Delta_{N}^{\nu^2}(x_1,\ldots,x_N)
\lim_{\eps\downarrow 0} \int_{-L/2}^{L/2} dy_1 e^{-ip_1y_1}
\cdots\int_{-L/2}^{L/2} dy_N e^{-ip_N y_N} \nonu\times
\prod_{j>j'}\check b(y_j-y_{j'};\eps_j+\eps_{j'})^{\nu^2} 
\prod_{j,\ell} \check b(y_j-x_\ell; 2\eps ) ^{-\nu^2} 
\nonueqaend
where $p_j=\tPiL n_j$ and
$
\check b(x,\eps):= \left(1 - e^{-\tPiL \eps} e^{i\tPiL x} \right) . 
$
Comparing with Eq.\ \Ref{Feta1} we see that
\nonueqa
\cP_{\eta_{\nu,N}(\vn) }(z_1,\ldots,z_N) = 
\lim_{\eps\downarrow 0} \int_{-L/2}^{L/2} dy_1 e^{-i\tPiL n_1y_1}
\cdots\int_{-L/2}^{L/2} dy_N e^{-i\tPiL n_N y_N} \nonu\times
\prod_{j>j'}\left( 1 - e^{-\tPiL \eps} e^{ i\tPiL (y_j-y_{j'} )} 
\right)^{\nu^2}  
\prod_{j,\ell} \left( 1 - e^{-\tPiL \eps}  e^{i\tPiL y_j} 
z_\ell
\right)^{-\nu^2} . 
\nonueqaend
We now can expand the integrand in a Taylor series in the exponentials 
and then perform the $y_j$-integrations. The final result is
\eqa
\label{polynom}
\cP_{\eta_{\nu,N}(\vn) }(z_1,\ldots,z_N) = 
 L^N \sum{}\!'
\prod_{j=1}^N \prod_{j'=1}^{j-1} \prod_{\ell=1}^N 
\binom{\nu^2}{\mu_{jj'} } \binom{-\nu^2}{ m_{j\ell} } 
(-1)^{\mu_{jj'} } (-z_\ell)^{m_{j\ell} }
\eqaend
where $\binom{\pm\nu^2}{n}$ are the binomial coefficients as usual and 
$\sum{}\!'$ here means summation over all $\mu_{jj'},
m_{j\ell} \in\N_0$ such that 
\eq
\sum_{j'=1}^{j-1} \mu_{jj'} - \sum_{j'=j+1}^N \mu_{j'j} + 
\sum_{\ell=1}^N m_{j\ell} = n_j \quad \mbox{ for 
$j=1,2\ldots N$ .} 
\eqend

\subsection{The braid group}

The braid group will not play a role in our deliberations however we
mention one observation for completeness.
We define operators on the $N$-anyon subspace
as follows. On a vector
$$\phi_\eps^{\nu}(x_1)\cdots \phi_\eps^{\nu}(x_N) 
\Omega$$
define, for $i\in \{1,2,...,N-1\}$,
$\sigma_i$ to be the operator which
 interchanges the $i^{th}$ and $(i+1)^{th}$ arguments and multiplies
by the phase: 
$$e^{-i \pi \nu^2 \sgn(x_i-x_{i+1};\eps)/2}.$$
An easy calculation reveals that the braid relations hold:
$$\sigma_i \sigma _j =\sigma_j\sigma _i,\ \  |i-j|>1,$$
$$\sigma_i^2 =1$$
$$\sigma_j \sigma_i \sigma_j =\sigma_i \sigma_j \sigma_i,\ \  |i-j|=1.$$
So we have a braid group action on each $N$-anyon subspace.

\subsection*{Appendix C. Proofs}

\noindent{\bf C.1 Proof of Proposition 1:}
According to Eqs.\ \Ref{phinup} we have to compute 
$$
(\cdot):=
\int_{-L/2}^{L/2}dy\, e^{ipy}
\check{\phi}_\eps^\nu(y)\eta_b,$$
(we used \Ref{phinup2}) for $\eta_b$ as in Eq.\ \Ref{etab}, and show 
that this has a well-defined strong limit $\eps\downarrow 0$ which is in 
$\cD_b$.  We note that Eq.\ \Ref{inv} implies for all $q\in\Lambda^*$ 
\eq
\label{caa}
\phi_\eps^\nu(x)\b(q)=\left[\b(q)-\nu e^{-iq x - |q|\eps}I \right]
\phi_\eps^\nu(x)
\eqend
(we used Eqs.\  \Ref{bosons}, \Ref{anyon} and 
$S(\tilde f_{y,\eps},\epsilon_{q}) = e^{-iq y-|q|\eps}$), and similarly for 
$\check{\phi}$. We thus obtain 
\nonueqa
\check{\phi}_\eps^\nu(y)\eta_b = 
\left[\b(-q_1)-\nu e^{iq_1(y+i\eps)} I \right] \cdots \left[\b(-q_n)-\nu 
e^{iq_n(y+i\eps)} I \right] \nonu\times \sum_{m=0}^\infty \frac{(i\nu)^m}{m!}
d\Gamma(\alpha^+_{y,\eps} )^m R^{\ell+\nu/\nu_0}\Omega
\nonueqaend
where we used Eqs.\ \Ref{phinup2} and \Ref{imp}.  Now 
$$ 
d\Gamma(\alpha^+_{y,\eps}) = -i\sum_{j=1}^\infty \frac{1}{ j}\b(-\tPiL j) 
e^{-i\tPiL j(y-i\eps)}, $$ thus
\nonueqa
(\cdot)= \int_{-L/2}^{L/2} dy \, e^{ipy} \left[\b(-q_1)-\nu 
e^{iq_1(y+i\eps)} I \right] \cdots \left[\b(-q_n)-\nu e^{iq_n(y+i\eps)}I
\right] \nonu\times
\sum_{m_1,m_2,\ldots=0}^\infty \prod_{j=1}^\infty 
\frac{\nu^{m_j} }{m_j! j^{m_j} }e^{-i\tPiL j(y-i\eps) m_j} 
\b(-\tPiL j)^{m_j} R^{\ell+\nu/\nu_0}\Omega .
\nonueqaend
We see that only terms with
\eq
\label{restr}
\frac{2\pi}{L} 
\sum_{j=1}^\infty  j m_j = p+\delta_1 q_1+\cdots +\delta_n q_n,
\quad \delta_i=0,1, m_j=0,1,2,\ldots 
\eqend
are non-zero after the integration, and this is only a {\em finite number 
of terms}. Notice that the
$\eps$ dependence arises only in the 
scalars multiplying these finitely many vectors.
 It is now obvious that the limit $\hat\phi^\nu(p)\eta_b = 
\lim_{\eps\downarrow 0} (\cdot)$ exists in norm, and we obtain
\eqa
\hat\phi^\nu(p)\eta_b =
L\sum{}\!'
(-\nu)^{\delta_1+\ldots+\delta_n} \b(-q_1)^{1-\delta_1}\cdots 
\b(-q_n)^{1-\delta_n}
\prod{}\!'_j
\frac{\nu^{m_j} }{m_j! j^{m_j} } \b(-\tPiL j)^{m_j} R^{\ell+\nu/\nu_0}\Omega
\eqaend
where $\sum{}\!'$ means that the sum is
 over all $\delta_i$ and 
$m_j$ obeying the condition Eq.\ \Ref{restr}, and $\prod{}\!'_j$
indicates that the product over $j$ is also constrained by \Ref{restr}.
This is manifestly a vector in $\cD_b$. 

Especially for $p=0$ and $n=0$ we get Eq.\ \Ref{spec}. \QED

\noindent{\bf C.2 Proof of Proposition 2:}
We compute 
$(\cdot):=\left<\eta_b, 
\phi_\eps^{\nu}(y_1)\cdots \phi_\eps^{\nu}(y_N) 
\Omega\right>$
with $\eta_b$ Eq.\ \Ref{etab}. We obtain 
$$
(\cdot) = \left<\Omega, R^{-\ell}\b(q_n)\cdots\b(q_1) 
\phi_\eps^{\nu}(y_1)\cdots \phi_\eps^{\nu}(y_N) 
\Omega\right>,
$$
and by a simple computation,
\nonueqa
(\cdot) = \left( \nu e^{-iq_1 (y_1-i\eps) }+ \cdots \nu e^{-iq_1 
(y_N-i\eps)} \right) \left<\Omega, R^{-\ell}\b(q_n)\cdots\b(q_2) 
\phi_\eps^{\nu}(y_1)\cdots + \phi_\eps^{\nu}(y_N) 
\Omega\right> \nonu
= \ldots = 
\prod_{j=1}^n\left( \nu e^{-iq_j (y_1-i\eps) }+ \cdots \nu e^{-iq_j 
(y_N-i\eps)} \right)\left<\Omega, R^{-\ell}
\phi_\eps^{\nu}(y_1)\cdots \phi_\eps^{\nu}(y_N) \Omega\right> \\
= \prod_{j=1}^n\left( \sum_{k=1}^N \nu e^{-iq_j (y_k-i\eps) } \right) 
\delta_{\ell,N\nu/\nu_0} e^{-i\pi\nu^2(y_1+\ldots +y_N)N/L 
}\prod_{j<k}b(y_j-y_k;2\eps)^{\nu^2} \nonueqaend with $b$ defined in 
Eq.\ \Ref{b} (we used Eqs.\ \Ref{caa} and \Ref{bOm} in the first two 
lines and Eqs.\ \Ref{corr}--\Ref{corr1} in the third).  It is now 
manifest that the limit $F^\nu_{\eta_b}=\lim_{\eps\downarrow 0}(\cdot)$ 
exists, and we obtain Eq.\ \Ref{Fetab} which is obviously in 
$L^2(S_L^N)$.  \QED

\noindent{\bf C.3 Proof of Proposition 3:}
Using Eqs.\ \Ref{phinup2} and \Ref{noo} we obtain by a straightforward 
computation
\nonueqa
\eta_{\nu,N}(\vn)= \lim_{\eps_1,\ldots,\eps_m\downarrow 0}
\int_{-L/2}^{L/2} dy_1 e^{2\pi i n_1 y_1/L}
\cdots \int_{-L/2}^{L/2} dy_N e^{2\pi i n_N y_N/L} (\cdots)
\nonueqaend
where $(\cdots)$ equals 
\nonueqa
\prod_{j<\ell} \left( 1-e^{2\pi i (y_j 
-y_\ell)/L-2\pi(\eps_j+\eps_\ell) } \right)^{\nu^2}
\exp\left(\nu\sum_{k=1}^\infty \frac{1}{k}\b(-\tPiL 
k)\left[\sum_{k'=1}^N 
e^{-2\pi i k (y_{k'}-i\eps_{k'})/L}\right] \right) R^{N}\Omega . 
\nonueqaend
(Note that the limit is in the strong sense.)
Expanding the latter in powers of $e^{i y_{j}/L}$ shows that this is 
a sum of terms proportional to 
$$
\prod_{j<\ell} e^{i q_{j\ell}(y_j- y_{\ell} + 
i\eps_j- i\eps_{\ell}  )} \prod_{k} e^{ -iq_{k} ( y_{k} - i\eps_{k} )  }
$$
where $q_{j\ell}$ and $q_{k}$ are in $\Lambda^*$ and {\em non-negative}. Thus 
$\eta_{\nu,N}(\vn)$ can be non-zero only if 
\nonueqa
\frac{2\pi}{L} n_{\ell} - \sum_{j=1}^{\ell-1} q_{j\ell} 
+ \sum_{j=\ell+1}^{N} q_{\ell j} - q_{\ell} = 0 \quad
\mbox{ for $\ell = 1,2,\ldots N$}
\nonueqaend
for at least one set of non-negative numbers 
$q_{j\ell}, q_{k}\in\Lambda^* $. 
Adding these conditions we get
$\sum_{\ell =1}^N n_\ell -\sum_{\ell=1}^N q_\ell =0$ which implies 
Eq.\ \Ref{conda}. Moreover, if these conditions hold then
$$
q_{j\ell} = \frac{2\pi}{L} n_{\ell} + \sum_{\ell'=
\ell+1}^{N} q_{\ell \ell'} -(\geq 0) \quad 
\forall j<\ell
$$
with `$(\geq 0)$' terms which always are non-negative.  By induction we 
obtain from this \nonueqa q_{j\ell} = \frac{2\pi}{L} n_{\ell} + 
\sum_{j'=\ell+1}^k 
2^{j'-1-\ell}\left( \frac{2\pi}{L} n_{j'} + 
\sum_{\ell'=j'+1}^{N} q_{j' \ell'}\right) 
-(\geq 0) \quad \forall j<\ell \nonueqaend which should be 
positive. Setting $k=N$ this implies Eq.\ \Ref{cond}.  \QED

\section{$W$-charges}

\subsection{Motivation} 
There are self-adjoint operators $W^{s}$ on $\cF$ obeying
\eq
\label{obey}
[W^{s}, \hat\psi^{*}(k)]= k^{s-1}\hat\psi^{*}(k)\quad \forall 
k\in\Lambda_0^*, \quad 
W^{s}\Omega =0\qquad (s\in\N) .   
\eqend
If we introduce an operator valued distribution $\psi^*(x)$ such that 
\[
\hat \psi^*(k)= \psi^*(e_k)=\int_{S_L}dx\, 
e_k(x)\psi^*(x),
\]
the commutator relations in Eq.\ \Ref{obey} are (formally\footnote{our 
results below will actually give a precise mathematical meaning to 
this}) equivalent to 
\eq
\label{formal}
[W^{s},\psi^*(x)] = i^{s-1}\frac{\partial^{s-1} }{\partial x^{s-1} 
}\psi^*(x) . \eqend  
These operators $W^s$ can be represented in terms of the 
operators $\b(p)$ Eq.\ \Ref{bosons}, \eqa
\label{fermionW}
W^1&=&\b(0) \nonu
W^2 &=& \frac{\pi}{L}\sum_{p\in\Lambda^*}
\xxa \b(p) \b(-p) \xxe  \nonu
W^3 &=& \frac{4\pi^2}{3L^2} \sum_{p_1,p_2\in\Lambda^*} 
\xxa \b(p_1) \b(p_2)\b(-p_1-p_2) \xxe - 
\frac{\pi^2}{3 L^2} \b(0) \\
&& \mbox{etc.} \nonumber \eqaend These formulas are known in the 
physics literature (see e.g.\ \cite{Ba}). We shall construct operators 
which obey similar relations with the anyon field operators 
$\phi_\eps^\nu(x)$.  To explain our method, we will first present a 
construction of operators $W^s$ obeying Eq.\ \Ref{obey} 
for all $s\in\N$.  We then show how to partly extend this to anyons.  
The extension is essentially trivial for $s=1,2$.  The first 
non-trivial case is $s=3$.  We propose a natural generalization of 
$W^3$ and show that it corresponds to a `second quantization' of the 
CS Hamiltonian Eq.\ \Ref{Sutherland}, as described in the 
Introduction.

To simplify our notation, we set $\nu_0=\nu$ in the rest of the paper. 

\subsection{$W$-charges for fermions}
We define 
\eq
\label{cWy}
\cW^\nu_{\eps}(y;a) : = 
N^\nu(a) \left( \xxa e^{i\nu d\Gamma(\tilde f_{y+a,\eps}- \tilde  
f_{y,\eps}) } \xxe - I \right)  , 
\eqend 
with functions $\tilde f_{y,\eps}$ given by equations \Ref{tildef}, 
\Ref{alpha} and the normalization constant
\eq
\label{Na}
N^\nu(a) =\frac{i}{ 2 L \nu^2\cos^{\nu^2}(\PiL a)\tan(\PiL a) }.
\eqend
In this Section we are mainly interested in the fermion case where 
$\nu=1$, but in our discussion on anyons later we will need these 
formulas for general non-zero $\nu\in\R$. 

We claim that
\eq
\label{res1}
\cW^\nu(a) := \lim_{\eps\downarrow 0}\int_{-L/2}^{L/2} dy\,
\cW^\nu_{\eps}(y;a) = 
\sum_{s=1}^\infty \frac{ (-ia)^{s-1} \nu^{s-2} }{ (s-1) !} W^{\nu,s}
\eqend 
defines an operator valued generating function for
operators $W^{\nu,s}$, $s\in\N$. To be more precise:

\vspace*{0.45cm} {\bf \noindent Lemma 2:} {\em For all $a\in \R$ and 
non-zero $\nu\in\R$, the operators $\cW^\nu(a)$ Eqs.\ 
\Ref{cWy}--\Ref{res1} are well-defined on $\cD_b$ and leave $\cD_b$ 
invariant.  Especially, \eq
\label{WOm}
\cW^\nu(a)\Omega = 0 .  \eqend Moreover, Eq.\ \Ref{res1} defines a family
of 
operators $W^{\nu,s}$, $s\in\N$, which have $\cD_b$ as a common, 
dense invariant domain of definition.} 
\vspace*{0.45cm}

The proof of this result is in Appendix D. We now show how to compute 
these operators  $W^{\nu,s}$ explicitly. 
We define
\eq
\tilde\delta_{y,\eps}(x):=-\frac{1}{2\pi }\partial_y \tilde f_{y,\eps}(x)  = 
\delta_{y,\eps}(x) + \frac{ (1-\nu) }{L} 
\eqend 
where $\partial_y=\frac{\partial}{\partial y}$, and\footnote{Note that 
$\b(p)=\lim_{\eps\downarrow 0}\int_{-L/2}^{L/2}dx\, \rho_\eps(x) 
e^{-ipx} $, which 
motivates our notation.}
\eq
\tilde\rho_\eps(y): = d\Gamma(\tilde\delta_{y,\eps}) = 
\rho_\eps(y) + \frac{ (1-\nu) }{L} Q.
\eqend
With that we obtain 
$$
d\Gamma(\tilde f_{y+a,\eps}- \tilde f_{y,\eps})  = -2\pi \sum_{k=1}^\infty 
\frac{a^s}{s!} \partial_y^{s-1}\tilde\rho_\eps(y) ,  
$$
and one can expand $\cW^\nu_{\eps}(y;a) $ Eq.\ 
\Ref{res1} in a formal power series in $a$. A straightforward computation 
then gives
\eqa
\label{res2}
W^{\nu,1} &=& \int_{-L/2}^{L/2} dy\,\xxa 
\tilde\rho_\eps(y)\left.\xxe\right|_{\eps\downarrow 0} 
\nonu W^{\nu,2} 
&=& \pi \int_{-L/2}^{L/2} dy\, \xxa \tilde\rho_\eps(y)^2 
\left.\xxe\right|_{\eps\downarrow 0} \nonu W^{\nu,3} 
&=&\frac{4\pi^2}{3}\int_{-L/2}^{L/2} dy\, \xxa 
\tilde\rho_\eps(y)^3\left.\xxe\right|_{\eps\downarrow 0} + 
\frac{\pi^2}{3L^2\nu^2}(2-3\nu^2) W^{\nu,1} \\
&& \mbox{etc.} \nonumber \eqaend (this list can be easily extended 
with the help of a symbolic programming language like MAPLE).  Note 
that for $\nu=1$, these are identical to the operators in Eq.\ 
\Ref{fermionW}, $ W^{1,s}=W^s $ for $s=1,2,3$.  Later we will also 
need the following formulas which are obtained by simple computations 
from the definitions above, \eqa
\label{res21}
W^{\nu,1} &=&  (2-\nu)Q \nonu 
W^{\nu,2} &=&  W^{2}+ \frac{\pi}{L}  (1-\nu)(1-3\nu)Q^2\nonu 
W^{\nu,3} &=&  W^{3} + 4\frac{\pi}{L} (1-\nu)Q W^{2} +
\frac{4}{3}\left(\frac{\pi}{L}\right)^2 (1-\nu)^2(4-\nu)Q^3
\nonu &&
- \frac{2}{3}\left(\frac{\pi}{L}\right)^2 (1-\nu)(1-\nu-3\nu^2)Q . 
\\
&& \mbox{etc.} \nonumber
\eqaend

The result described in the last 
subsection can now be stated as follows. 

\vspace*{0.45cm} 
{\bf \noindent  Theorem 1:} {\em The operators $W^{1,s}$ obey the 
relations Eq.\ \Ref{obey} i.e.\ $W^{1,s}=W^s$ for all $s\in\N$.}
\vspace*{0.45cm} 

{\em \noindent Proof:} We recall Eq.\ \Ref{WOm} for $\nu=1$. Here 
we will show that
\eq
\label{crucial}
[\cW^1(a) , \hat\psi^*(k)] = e^{-ika} \hat\psi^*(k) \eqend These two 
relations prove the result, as can be seen by an expansion
in a formal power series in $a$ and using Eq.\ \Ref{res1}.

To prove Eq.\ \Ref{crucial} we use the boson-fermion correspondence 
Eq.\ \Ref{bosonfermion}.  We thus compute the commutator of 
$\cW^1_{\eps'} (y)$ with $\phi^1_\eps(x)=\Gamma(e^{i
f_{x,\eps}})$.  With Eqs.\
\Ref{noo}, \Ref{tildeS} and \Ref{blips} we obtain $$ 
[\cW^1_{\eps'}(y;a), \phi^1_\eps(x)] = (\cdots)
\xxa\Gamma(e^{i[f_{x,\eps}+ f_{y+a,\eps'}-f_{y,\eps'} ]})\xxe $$ 
with
$$ (\cdots):= 
N^1(a)\left( \frac{\sin\PiL(y+a-x+i\tilde\eps) }{\sin\PiL(y-x+i\tilde\eps) } 
-c.c.\right) = \frac{i}{2 L}\left(
\cot\PiL(y-x+i\tilde\eps)  - c.c. \right) 
$$
where $\tilde\eps=\eps+\eps'$ and $c.c.$ means the same term complex 
conjugated. We now use that
\eq
\label{ctg}
\pm\frac{i}{2 L} \cot\PiL(y-x\pm i\tilde\eps) = \frac{1}{2L} + 
\delta^\pm_{x,\tilde\eps}(y) \eqend which is easily seen by expanding 
the l.h.s as a Taylor series in $e^{\pm i(y-x)2\pi/L}e^{-\eps 
2\pi/L}$.  Thus $(\cdots)=\delta_{x,\tilde\eps}(y)$ independent of $a$ 
(!), and we obtain
$$
[\cW^1_{\eps'} (y;a),\phi^1_\eps(x)] = \delta_{x,\eps+\eps'}(y) 
\xxa\Gamma(e^{i[f_{x,\eps}+ f_{y+a,\eps'}-f_{y,\eps'} ]})\xxe. 
$$
Using Eqs.\ \Ref{res1} and \Ref{bosonfermion}  we thus obtain
for the l.h.s. of Eq.\ \Ref{crucial}, 
\nonueqa
\lim_{\eps\downarrow 0}\frac{1}{ \sqrt{2\pi L} } \int_{-L/2}^{L/2} dx 
\, e^{ikx} \lim_{\eps'\downarrow 0}\int_{-L/2}^{L/2} dy \, 
\delta_{x,\eps+\eps'}(y) \xxa\Gamma(e^{i[f_{x,\eps}+ 
f_{y+a,\eps'}-f_{y,\eps'} ]})\xxe \nonu = \lim_{\eps\downarrow 
0}\frac{1}{ \sqrt{2\pi L} } \int_{-L/2}^{L/2} dx \,e^{ikx} 
\xxa\Gamma(e^{if_{x+a,\eps}})\xxe \nonueqaend in the sense of strong 
convergence on a dense domain.  Recalling 
$\Gamma(e^{if_{x,\eps}})=\phi^1_\eps(x)$ and using Eq.\ 
\Ref{bosonfermion} again we obtain the r.h.s. of Eq.\ \Ref{crucial}.  
\QED

We finally discuss a technical point which will be important in the next 
Section: Our proof above shows that
$$
[\cW^1(a),\phi^1_\eps(x)] \simeq \phi^1_\eps(x+a)
$$
where `$\simeq$' means equality after smearing with appropriate test 
functions, and taking the strong
 limit $\eps\downarrow 0$ on an appropriate dense domain.  
It will be useful to characterize `$\simeq$' more explicitly as follows. 
Using Eq.\ \Ref{anyon} we define 
\eq\label{tildephi}
\tilde\phi^\nu_\eps(x;a):= 
\lim_{\eps'\downarrow 0}\int_{-L/2}^{L/2} dy \, 
\delta_{x,\eps+\eps'}(y) \xxa\Gamma(e^{i\nu[\tilde f_{x,\eps}+ \tilde 
f_{y+a,\eps'}-\tilde f_{y,\eps'} ]})\xxe \eqend
Then 
 $\tilde\phi^\nu_\eps(x;a)\simeq 
\phi^\nu_\eps(x+a) $. 
 We now define \eq \frac{\partial_\eps^{s-1} 
}{\partial_\eps x^{s-1} }\phi^\nu_\eps(x):= \left.  
\frac{\partial^{s-1} }{\partial x^{s-1} }\tilde\phi^\nu_\eps(x;a) 
\right|_{a=0} \eqend for $s=1,2,\ldots$, 
which we regard as $\eps$-deformed differentiations.  We specify the 
relation between these and the ordinary differentiations in the 
following

\vspace*{0.45cm} 
{\bf \noindent Lemma 3:} {\em 
\eq \frac{\partial_\eps^{s-1} 
}{\partial_\eps x^{s-1} }\phi^\nu_\eps(x) = \frac{\partial^{s-1} 
}{\partial x^{s-1} }\phi^\nu_\eps(x) + \eps\xxa 
c^{s,\nu}_\eps(x)\phi^\nu_\eps(x)\xxe \eqend where $c^{s,\nu}_\eps(x)$ 
is a well-defined operator-valued distribution
for $\eps\downarrow 0$.  Especially,\footnote{We will only need this 
for $s=1,2,3$ and 
thus do not specify the $c^{s,\nu}_\eps(x)$ for $s>3$.}
$$ c^{1,\nu}_\eps(x) = c^{2,\nu}_\eps(x) = 0,$$
\eqa 
\label{c123}
c^{3,\nu}_\eps(x) = \frac{(i \nu)^2 }{L^2}\sum_{p_1,p_2\in\Lambda^*} 
\hat\rho(p_1)\hat\rho(p_2) \, e^{i(p_1+p_2) x}
\frac{1}{\eps} \left(
e^{-\eps(|p_1+p_2|) } -e^{-\eps|p_1|-\eps|p_2| } 
\right) .
\eqaend
}
(The proof is a straightforward computation which we skip.)

\subsection{$W$-charges for anyons}
The considerations of the
preceding section may be extended to cover the case
of anyons i.e.\ $\nu$ an arbitrary non-zero real number. 
Using an argument similar to that in the proof of Theorem 1, we compute 
$$
[\cW^\nu_{\eps'}(y;a), \phi^\nu_\eps(x)] = (\cdots) 
\xxa\Gamma(e^{i\nu [\tilde f_{x,\eps}+ \tilde f_{y+a,\eps'}-
\tilde f_{y,\eps'} ]})\xxe 
$$
with $(\cdots)$ equal to 
\eqa
\label{cdots}
N^\nu (a) \left[ \left(\frac{\sin\PiL(y+a-x+i\tilde\eps) } 
{\sin\PiL(y-x+i\tilde\eps) }\right)^{\nu^2} -c.c.\right]\nonu = 
N^\nu(a)\cos^{\nu^2}(\PiL a)\left(1+\tanh(\PiL a) 
\cot\PiL(y-x+i\tilde\eps) \right)^{\nu^2} + c.c. \nonu 
\delta_{x,\tilde\eps}(y) -\half(\nu^2-1)a\partial_y 
\delta_{x,\tilde\eps}(y) + \cO(a^2) \eqaend where 
$\tilde\eps=\eps+\eps'$ (in the last line we Taylor expanded in $a$ 
and used $\cot^2(z)=-1-d\cot(z)/dz$ and Eq.\ \Ref{ctg}).  Integrating 
this in $y$, performing a partial integrations, and using Eq.\ 
\Ref{tildephi} we thus obtain
$$
[\cW^\nu(a), \phi^\nu_\eps(x)] = \tilde\phi^\nu_\eps(x;a)
+ i \pi \nu(\nu^2-1)a \xxa [\tilde\rho_\eps(x+a)-\tilde\rho_\eps(x)] 
\tilde\phi^\nu_\eps(x;a) \xxe +\cO(a^3) . 
$$ 
Comparing now equal powers of $a$ on both sides of the last equation 
using Eqs\ \Ref{tildephi}--\Ref{c123} we 
see that the generalization of Theorem 1 to anyons holds true only for 
$s=1,2$,
\eq
\label{that12}
[W^{\nu,s},\phi^\nu_\eps(x)] =\nu^{2-s} 
i^{s-1}\frac{\partial^{s-1}}{\partial x^{s-1}} \phi^\nu_\eps(x)\qquad  s=1,2
\eqend
but $s>2$ we get correction terms, e.g.\
\eq
\label{w3}
[W^{\nu,3},\phi^\nu_\eps(x)] = 
\frac{i^2}{\nu}\frac{\partial_\eps^2}{\partial_\eps 
x^2}\phi^\nu_\eps(x) + 2\pi i (\nu^2-1) \xxa
\tilde\rho_\eps(x)'\phi^\nu_\eps(x) \xxe 
\eqend
where $\tilde\rho_\eps(x)':=\partial_x\tilde\rho_\eps(x)$. 
We define
\eq
\cH^{\nu,1}:\; = \frac{1}{\nu} W^{\nu,1} ,\quad \cH^{\nu,2}:\; = W^{\nu,2} 
\eqend
which according to Eq.\ \Ref{that12} are the anyon $W$-charges 
for $s=1,2$. 

In the following we only consider the first non-trivial case $s=3$.
To proceed,  it is crucial to observe that  correction term in Eq.\ 
\Ref{w3} can be partly canceled using the following operator,
\eqa
\label{cC}
\cC = -\pi i\int_{-L/2}^{L/2} dy\, \xxa [\rho^+_\eps(y) - 
\rho^-_\eps(y)] \partial_y [\rho^+_\eps(y)+ \rho^-_\eps(y) 
]\left.\xxe\right|_{\eps\downarrow 0} \nonu = - \frac{2\pi}{L} 
\sum_{p>0} \xxa p\b(p)\b(-p) \xxe \eqaend where \eq
\label{rhopm}
\rho_{y,\eps}^\pm : = d\Gamma(\delta^\pm_{y,\eps}) . 
\eqend
This operator obeys the remarkable relations, 
\eq
\label{cc1}
\cC \phi^\nu_\eps(x) + \phi^\nu_\eps(x)\cC = 2\pi i \nu\xxa 
\tilde\rho_{\eps}(x)'\phi^\nu_\eps(x)\xxe + 2\xxa \cC 
\phi^\nu_\eps(x)\xxe  . \eqend 
The proof of this, 
which we now outline, is by a computation similar to the one leading to 
Eq.\ \Ref{w3}.  We consider the operator
\eqa
\label{V}
\cV_\eps(y;a,b) := \xxa     
e^{ -ia d\Gamma( i\delta^+_{y,\eps} -i\delta^-_{y,\eps}  ) +ib d\Gamma( 
\partial_y \delta^+_{y,\eps} + \partial_y \delta^-_{y,\eps}  )    }
\xxe
\eqaend
and observe that
\eq
\label{VC}
\cC = -\pi \lim_{\eps\downarrow 0} \left. 
\int_{-L/2}^{L/2} dx\, \frac{\partial}{\partial a} \frac{\partial}{\partial b}
\cV_\eps(y;a,b) \right|_{a=b=0} .   
\eqend
Using Eqs.\ \Ref{noo}, \Ref{tildeS} and \Ref{blips} 
one then computes
$$
 \cV_{\eps'}(y;a,b)\phi^\nu_\eps(x) + \phi^\nu_\eps(x)\cV_{\eps'}(y;a,b)
$$
which by a Taylor expansion in $a$ and $b$ and integrating in $y$
gives Eq.\ \Ref{cc1}. (For details see Appendix D, Proof of Lemma 4.)  

We also note that Eq.\ \Ref{bOm} implies \eq
\label{cCOm}
\cC R^\ell\Omega=0\quad \forall \ell\in\Z .  
\eqend 
Thus the operator 
\eq
\label{Hpmnu}
\cH^{\nu,3} = \nu W^{\nu,3} + (1-\nu^2) \cC
\eqend
obeys the relation 
$$
[\cH^{\nu,3},\phi^\nu_\eps(x)] = 
i^2\frac{\partial_\eps^2}{\partial_\eps x^2}\phi^\nu_\eps(x) +
2(1-\nu^2 ) (\xxa \cC 
\phi^\nu_\eps(x)\xxe - \phi^\nu_\eps(x)\cC ) . 
$$
Again there are correction terms, however, in contrast to the one in 
Eq.\ \Ref{w3} it vanishes when applied to vectors $R^w\Omega$! We 
obtain 
\eq
\label{that123}
[\cH^{\nu,3},\phi^\nu_\eps(x)]R^w\Omega = 
i^2\frac{\partial^2}{\partial x^2}
\phi^\nu_\eps(x)R^w\Omega 
\eqend
(we used Lemma 3 and $\xxa 
c^{s,\nu}_\eps(x)\phi^\nu_\eps(x)\xxe R^w\Omega=0$). 
This seems to be the best we can do to generalize the relation Eq.\ 
\Ref{formal} for $s=3$ to the anyon case.  

To fully appreciate this operator $\cH^{\nu,3}$ one has to extend the 
computation above to a product of multiple anyon operators.  We thus 
obtain our main result:

\vspace*{0.45cm} 
{\bf \noindent  Theorem 2}. {\em The operator
$\cH^{\nu,3}$ obeys the following relations, 
\eq
\label{that}
[\cH^{\nu,3}, \phi^\nu_{\eps_1}(x_1)\cdots  \phi^\nu_{\eps_N}(x_N)]R^w\Omega =
H_{N,\nu^2,\beps} \phi^\nu_{\eps_1}(x_1)\cdots  
\phi^\nu_{\eps_N}(x_N)R^w\Omega  
\eqend
for all integer $w$, where
\eq
\label{CM} 
H_{N,\nu^2,\beps} = -\sum_{k=1}^N \frac{\partial^2}{\partial x_k^2} 
+\sum_{ {\stackrel{k,\ell=1}{k\neq \ell} }}^N \frac{ 
(\PiL)^2\nu^2(\nu^2 - 1) }{\sin^2 
\PiL(x_k-x_\ell-i\sgn(k-\ell)(\eps_k+\eps_\ell) ) } + 
C_{N,\nu^2,\beps}(\vx) \eqend is a regularised version of the CS 
Hamiltonian Eq.\ \Ref{Sutherland}, i.e.\ the function 
$C_{N,\nu^2,\beps}(\vx)$\footnote{The 
interested reader can find the definition of this function Eq.\ 
\Ref{C} below.} is non-singular and vanishes uniformly as 
$\eps_j\downarrow 0$ for all $j=1,2.\ldots N$.
}
\vspace*{0.45cm}

The proof of this Theorem is a straightforward but tedious extension 
of the computation leading to Eq.\ \Ref{that123} (which is the special 
case $N=1$), and the interested reader can find it in Appendix D.

\subsection*{Appendix D: Proofs}

\subsubsection*{Proof of Lemma 2}
The argument here is very similar to the Proof of Proposition 1 and thus 
we can be brief.

For $\eta_b$ given by  equation \Ref{etab} we obtain
\nonueqa
\xxa \Gamma(e^{i\nu (\tilde f_{y+a,\eps}- \tilde 
f_{y,\eps}) }) \xxe \eta_b = 
\prod_{j=1}^n \left[\b(-q_j)-\nu e^{iq_j 
(y+a)-|q_j|\eps}I+ \nu e^{iq_jy-|q_j|\eps}I \right] \nonu\times
\sum_{m_1,m_2,\ldots=0}^\infty \prod_{j=1}^\infty 
\frac{\nu^{m_j} }{m_j! j^{m_j} }\left( e^{-ij\tPiL (y+a-i\eps)}-  
e^{-ij\tPiL (y-i\eps)}
\right)^{m_j} \b(-\tPiL j)^{m_j}
e^{-2\pi i\nu^2 a\ell /L} R^\ell\Omega . 
\nonueqaend
Just as in the proof of Proposition 1 in Appendix C
this shows that $$(\cdot):= \; \int_{-L/2}^{L/2} dy\; \xxa \Gamma(e^{i\nu 
(\tilde f_{y+a,\eps}- \tilde f_{y,\eps}) }) \eta_b$$ is a sum of a finite 
number of terms. As the $\eps$
dependence lies in the coefficients of 
this finite dimensional subspace the norm limit $\eps\downarrow 0$ 
exists and  is in 
$\cD_b$.  
Thus $\cW^\nu(a)\eta_b\in\cD_b$. Especially for 
$\eta_b=\Omega$, we obtain $(\cdot)= \Omega$, which implies Eq.\ 
\Ref{WOm}. \QED

\subsubsection*{D2. Proof of Theorem 2} We write
$$
\Phi^\nu_{\beps}(\vx)
: \; 
=\phi^\nu_{\eps_1}(x_1)\cdots  \phi^\nu_{\eps_N}(x_N) = 
\cJ^\nu_{\beps}(\vx) \xxa\Phi^\nu_{\beps}(\vx)\xxe 
$$
where 
$$\cJ^\nu_{\beps}(\vx) = 
\cJ^{\nu_1,\cdots,\nu_N}_{\eps_1,\cdots,\eps_N}(x_1,\ldots,x_N)
$$ 
are defined in Eq.\ \Ref{J}. We also use the short-hand notation
\eq
A*\Phi^\nu_{\beps}(\vx)
: = \; \cJ^\nu_{\beps}(\vx)
\xxa A \Phi^\nu_{\beps}(\vx)\xxe . 
\eqend
We compute
$$
[\cW^\nu(a), \Phi^\nu_{\beps}(\vx)] = \lim_{\eps'\downarrow 0} 
\int_{-L/2}^{L/2} dy\; (\cdots) 
\xxa\Gamma(e^{i\nu [\tilde f_{y+a,\eps'}- 
\tilde f_{y,\eps'} ]}) \xxe * \Phi^\nu_{\beps}(\vx) 
$$
with
\nonueqa 
(\cdots) = 
N^\nu (a) \left[ \prod_{j=1}^N 
\left(\frac{\sin\PiL(y+a-x_j+i\tilde\eps_j) } 
{\sin\PiL(y-x_j+i\tilde\eps_j) }\right)^{\nu^2} -c.c.\right] 
\\ = 
\sum_{\ell=1}^N (\cdot)_j 
N^\nu (a) \left[ \left(\frac{\sin\PiL(y+a-x_j+i\tilde\eps_j) } 
{\sin\PiL(y-x_j+i\tilde\eps_j) }\right)^{\nu^2} -c.c.\right]
\nonueqaend
where $\tilde\eps_j=\eps_j+\eps'$ and 
$$
(\cdot)_j = \prod_{\stackrel{k=1}{k\neq 
j}}^{N}
\left(\frac{\sin\PiL(y+a-x_\ell+i\sgn(j-k)\tilde\eps_\ell) } 
{\sin\PiL(y-x_\ell+i\sgn(j-k)\tilde\eps_\ell) }\right)^{\nu^2}  .
$$
Using Eq.\ \Ref{cdots} we obtain
\nonueqa
[\cW^\nu(a), \Phi^\nu_{\beps}(\vx) ] = \sum_{j=1}^N 
\lim_{\eps'\downarrow 0} 
\int_{-L/2}^{L/2} dy\; \delta_{x_j,\tilde\eps_j}(y) \\ 
\times 
\left[ 1 + \half(\nu^2-1)a \partial_y +\cO(a^2) \right](\cdot)_j
\xxa\Gamma(e^{i\nu [\tilde f_{y+a,\eps'}- 
\tilde f_{y,\eps'}]})\xxe * \Phi^\nu_{\beps}(\vx) . 
\nonueqaend

By a simple computation we see that this equals
\nonueqa
\sum_{j=1}^N \left(\tilde\Phi^\nu_{\beps}(\vx;a\ve_j)  + i\pi\nu (\nu^2-1)a 
[\tilde\rho_{\eps_j}(y_j+a)-
\tilde\rho_{\eps_j}(y_j)]*\tilde\Phi^\nu_{\beps}(\vx;a\ve_j)   
\right)
\nonu +  
\sum_{\stackrel{j,k=1}{j\neq 
k}}^N \nu^2(\nu^2-1) \PiL \left[\cot\PiL (y_{kj}+a) - \cot\PiL (y_{kj}) 
\right] \tilde\Phi^\nu_{\beps}(\vx;a\ve_j) +\cO(a^3) . 
\nonueqaend
where $y_{jk}=y_j-y_k + i \sgn(k-j)(\eps_j+\eps_k)$ and
$$
\tilde\Phi^\nu_{\beps}(\vx;a\ve_j) : \; 
=\phi^\nu_{\eps_1}(x_1)\cdots 
\tilde\phi^\nu_{\eps_j}(x_j;a)
\cdots  \phi^\nu_{\eps_N}(x_N)  
$$
with $\tilde\phi^\nu_{\eps}(x;a)$ defined in Eq.\ \Ref{tildephi}.  
Collecting the terms proportional to $a^2$ on both sides of this 
equation we obtain, \eqa {[}W^{\nu,3}, \Phi^\nu_{\beps}(\vx){]} &=& 
\frac{1}{\nu} \tilde H_{N,\nu^2,\beps} \Phi^\nu_{\beps}(\vx) + \sum_{j=1}^N
2\pi i(\nu^2-1) \tilde\rho'_{\eps_j}(y_j)*\Phi^\nu_{\beps}(\vx)    
\eqaend
with 
\eq
\label{CMtilde} 
\tilde H_{N,\nu^2,\beps} = -\sum_{k=1}^N 
\frac{\partial_\eps^2}{\partial_\eps x_k^2}
+\sum_{ {\stackrel{k,\ell=1}{k\neq \ell} }}^N
\frac{ (\PiL)^2\nu^2(\nu^2 - 1) }{\sin^2 
\PiL(x_k-x_\ell-i\sgn(k-\ell)(\eps_k+\eps_\ell) ) } 
\eqend
and $\partial_\eps^2/\partial_\eps x_j^2$ as characterized in Lemma 3. 
To proceed, we need to generalize Eq.\ \Ref{cc1}: 

\vspace*{0.45cm} 
{\bf \noindent  Lemma 4:} {\em
The operator $\cC$ given in Eq.\ \Ref{cC} satisfies the following relations
\eq
\label{cc}
\cC \Phi^\nu_{\beps}(\vx) + \Phi^\nu_{\beps}(\vx)\cC = 
2\xx \cC \Phi^\nu_{\beps}(\vx) \xx + \sum_{j=1}^N  
2\pi i \nu \tilde\rho'_{\eps_j}(y_j)*\Phi^\nu_{\beps}(\vx) . 
\eqend
}
\vspace*{0.45cm} 

\noindent Thus with $\cH^{\nu,3}$ Eq.\ \Ref{Hpmnu}, \eq 
[\cH^{\nu,3},\Phi^\nu_{\beps}(\vx)] = \tilde H_{N,\nu^2,\beps} 
\Phi^\nu_{\beps}(\vx) + 2 (1-\nu^2) (\xx \Phi^\nu_{\beps}(\vx)\cC \xx 
- \Phi^\nu_{\beps}(\vx)\cC) .  \eqend Applying this equation to the 
state $R^\ell\Omega$, eq.\ \Ref{cCOm} implies that only the second 
term on the r.h.s. vanishes, and we obtain Eq.\ \Ref{that} where we 
set \eq
\label{C}
C_{N,\nu^2,\beps}(\vx)\Phi^\nu_{\beps}(\vx)R^\ell\Omega : \; 
=-\sum_{j=1}^N \eps_j \phi^\nu_{\eps_1}(x_1)\cdots \xxa 
c^{3,\nu}_{\eps_j}(x_j)\phi^\nu_{\eps_j}(x_j)\xxe \cdots 
\phi^\nu_{\eps_N}(x_N)R^\ell\Omega .  \eqend It is easy to see that 
this defines a function $C_{N,\nu^2,\beps}(\vx)$ which can be 
calculated explicitly, however, we only need that this function is 
non-singular and vanishes uniformly as $\eps\downarrow0$ for all $j$, 
which is obvious (see Eqs.\ \Ref{c123} and \Ref{caa}).

\QED

\subsubsection*{D3. Proof of Lemma 4} 
We consider the operator $\cV_\eps(y;a,b)$ Eq.\ \Ref{V} and recall 
Eq.\ \Ref{VC}. Using Eqs.\ \Ref{noo}, \Ref{tildeS} and \Ref{blips} 
we compute
 \eq
 \label{ab}
 (\star):= 
 \cV_{\eps'}(y;a,b)\Phi^\nu_{\beps}(\vx) + \Phi^\nu_{\beps}(\vx)\cV_{\eps'}(y;a,b)
 = (\cdots)_2 \xxa \cV_{\eps'}(y;a,b)\Phi^\nu_{\beps}(\vx)\xxe
 \eqend
where
\nonueqa
(\cdots)_2= e^{\nu(-a+ib\partial_y)
\sum_{j=1}^N\delta^{+}_{x_j,\eps_j+\eps'}(y) } + c.c. 
\\
=   1 -  \nu a \sum_{j=1}^N[\delta^{+}_{x_j,\eps_j+\eps'}(y) +
\delta^{-}_{x_j,\eps_j+\eps'}(y)] +
\sum_{j=1}^Ni\nu b\partial_y[\delta^{+}_{x_j,\eps_j+\eps'}(y) - 
\delta^{-}_{x_j,\eps_j+\eps'}(y)] \\
- \frac{i\nu^2}{2} ab \sum_{j=1}^N\partial_y[ 
\delta^{+}_{x_j,\eps_j+\eps'}(y)^2 - 
\delta^{-}_{x_j,\eps_j+\eps'}(y)^2] + \cO(a^2)+\cO(b^2) .  \nonueqaend 
Moreover, using Eq.\ \Ref{rhopm} we expand \nonueqa \cV_\eps(y;a,b) = 
I + a\xxa[\rho^+_\eps(y) - \rho^-_\eps(y) ] \xxe + i b\partial_y 
\xxa[\rho^+_\eps(y) + \rho^-_\eps(y) ] \xxe \\
+iab\xxa[\rho^+_\eps(y) - 
\rho^-_\eps(y)]\partial_y[\rho^+_\eps(y) + \rho^-_\eps(y)]\xxe
+ \cO(a^2)+\cO(b^2) .  
\nonueqaend
Using the relation
\eq
\int_{-L/2}^{L_2} dy\, \delta^\sigma_{x,\eps+\eps'}(y)
\rho_{\eps'}^{\sigma'}(y)= 
\delta_{\sigma\sigma'}\rho_{\eps+2\eps'}^{\sigma'}(x), \quad 
\sigma,\sigma'=\pm 
\eqend
and
\eq
\int_{-L/2}^{L_2} dy\, \rho_{\eps'}^{\pm}(y) \rho_{\eps'}^{\pm}(y) = 0 
\eqend
we may calculate 
$$
-\pi \lim_{\eps\downarrow 0} \left. 
\int_{-L/2}^{L/2} dx\, \frac{\partial}{\partial a} \frac{\partial}{\partial b}
(\star) \right|_{a=b=0}, 
$$
and obtain Eq.\ \Ref{cc}. \QED

\section{The Calogero-Sutherland Hamiltonian and its eigenfunctions}
\label{suther}
We are now ready to
show how the results of the last Section provide the means to 
construct eigenfunctions and the corresponding eigenvalues of the 
CS Hamiltonian Eq.\ \Ref{Sutherland}. 

\subsection{Eigenfunctions from anyon correlation functions}
We claim that Theorem 2 essentially relates these eigenfunctions of 
the Sutherland Hamiltonian 
$H_{N,\nu^2}$ Eq.\ \Ref{Sutherland}, to the eigenvectors of the 
operator $\cH^{\nu,3}$.  In fact the key step is just to observe the 
following elementary corollary of Theorem 2.

{\bf \noindent Proposition 4}. 
{\em Let $\eta\in\cD_b$.  Then 
\eq \lim_{\eps\downarrow 0} 
<\eta, \cH^{\nu,3} \phi^\nu_\eps(x_1)\cdots \phi^\nu_\eps(x_N) \Omega> 
= H_{N,\nu^2} F^\nu_\eta(x_1,\ldots,x_N)  \eqend 
where $F^\nu_\eta$ is defined in Eq.\ \Ref{Feta}
Especially, if $\eta$ 
is an eigenvector of $\cH^{\nu,3}$ with the eigenvalue $\cE$ then 
$F^\nu_\eta$ is an eigenfunction of $H_{N,\nu^2}$ with the same eigenvalue 
$\cE$.  } \vspace*{0.45cm}

The immediate next question is to ask if our method constructs all 
eigenvectors of \Ref{Sutherland}.  We answer this in two steps.  We 
first state and prove another consequence of Theorem 2.

\vspace*{0.45cm}
{\bf \noindent Proposition 5}. {\em The  vectors $\eta_{\nu,N}(\vn)$ 
defined in Eq.\ \Ref{eta} are in $\cD_b$, and 
they obey
\eq
\label{c2}
\cH^{\nu,3}\eta_{\nu,N}(\vn)= 
\cE_{\nu,N}(\vn)\eta_{\nu,N}(\vn) + 
\gamma \sum_{j<\ell}\sum_{n=1}^\infty n 
\eta_{\nu,N}(\vn + n[ \ve_j-\ve_\ell ]) 
\eqend
with
\eq
\label{E}
\cE_{\nu,N}(\vn) = \sum_{j=1}^m P_{j}^2 ,
\eqend
$P_{j}$ defined in Eq.\ \Ref{Pj}, 
\eq
\label{gamma}
\gamma:= \; 2\nu^2(\nu^2-1) \left( \frac{2\pi}{L}\right )^2 , 
\eqend
and 
$\ve_1=(1,0,\ldots,0)$, $\ve_2=(0,1,\ldots,0)$, \ldots, $\ve_m=(0,0,\ldots,1)$. 
}
\vspace*{0.45cm}

\noindent {\em Proof:} 
We use Eqs.\ \Ref{eta}, \Ref{phinup} and Theorem 2 to write 
$$
\cH^{\nu,3}\eta_{\nu,N}(\vn)= (\cdot)_1 + (\cdot)_2 + (\cdot)_3 
$$
with
\nonueqa
(\cdot)_1 = \sum_{j=1}^N 
\lim_{\eps_1,\ldots,\eps_N\downarrow 0}\int_{-L/2}^{L/2} dx_1 e^{iP_1x_1}
\cdots \int_{-L/2}^{L/2} dx_N e^{iP_N x_N}\nonu\times
\left( -\frac{\partial^2}{\partial x_j^2} \right) 
\phi^{\nu}_{\eps_1}(x_1) \cdots \phi^{\nu}_{\eps_N}(x_N) 
\Omega ,
\nonueqaend
\nonueqa
(\cdot)_2 = 
\lim_{\eps_1,\ldots,\eps_N\downarrow 0}\int_{-L/2}^{L/2} dx_1 e^{ip_1x_1}
\cdots \int_{-L/2}^{L/2} dx_N e^{ip_N x_N}\nonu\times 
C_{N,\nu^2,\beps}(x_1,\ldots, x_N)
\check{\phi}^{\nu}_{\eps_1}(x_1) \cdots 
\check{\phi}^{\nu}_{\eps_N}(x_N) 
\Omega
\nonueqaend
and
\nonueqa
(\cdot)_3 = \sum_{j<\ell} \gamma 
\lim_{\eps_1,\ldots,\eps_N\downarrow 0}\int_{-L/2}^{L/2} dx_1 e^{ip_1x_1}
\cdots \int_{-L/2}^{L/2} dx_N e^{ip_N x_N}\nonu\times
\cS(x_j-x_\ell;\eps_k+\eps_\ell)
\check{\phi}^{\nu}_{\eps_1}(x_1) \cdots 
\check{\phi}^{\nu}_{\eps_N}(x_N) 
\Omega
\nonueqaend
where $\gamma$ is defined in Eq.\ \Ref{gamma} and 
$$
\cS(x;\eps) = \frac{1}{2 \sin^2 ( \PiL(r+i\eps) ) } = - 2\sum_{n=1}^\infty 
n e^{2\pi i n(r+i\eps)/L}   
$$
(the last equality is obtained by a Taylor expansion in $e^{2\pi i 
(r+i\eps)/L}$). 
Recalling Eq.\ \Ref{etanuN}, a simple computation implies
$$
(\cdot)_3 = \gamma \sum_{j<\ell}\sum_{n=1}^\infty n 
\eta_{\nu,N}(\vn + n[ \ve_j-\ve_\ell ]) . 
$$
Moreover, using Eq.\ \Ref{C} we see that $(\cdot)_2=0$. 
The remaining term is easily computed by partial integrations, 
$$
(\cdot)_1 = \sum_{j=1}^N P_{j}^2\eta_{\nu,N}(\vn) .
$$
This gives Eqs.\ \Ref{c2}--\Ref{E}. \QED
\vspace*{0.45cm}

Based on this result we can now present a simple algorithm to construct 
eigenvectors of the operator $\cH^{\nu,3}$. 
For that we find it convenient to use the following notation. For 
$\vmu\in\N_0^{N(N-1)/2}$ we write
\eq
\label{notation}
\vmu =(\mu_{j\ell})_{1\leq j<\ell\leq N}
= \sum_{j=1}^N\sum_{\ell=j+1}^N \mu_{j\ell} \vE_{j\ell},\quad 
\mu_{j\ell}\in \N_0
\eqend
which defines a canonical basis $(\vE_{j\ell})_{1\leq j<\ell\leq N}$  
in $\N_0^{N(N-1)/2}$. Moreover, we write
\newcommand{\ppls}{ \dot{+} }
\newcommand{\pmls}{ \dot{\pm} }
\newcommand{\mmls}{ \dot{-} }
\eq
\vn\pmls \vmu := \vn \pm \sum_{j<\ell} \mu_{j\ell}[\ve_j- \ve_\ell]. 
\eqend
E.g.,  $\vn\ppls n\vE_{j\ell}=\vn + n[ \ve_j-\ve_\ell ]$. 
We also write ${\bf 0}$ for the zero element 
in $\N_0^{N(N-1)/2}$, i.e.\ $\vn\pmls {\bf 0} =\vn$. 

Proposition 4 suggests the ansatz
\eq
\label{1}
\Psi = \sum_{ \vmu\in\N_0^{N(N-1)/2} } 
\alpha(\vmu ) \eta_{\nu,N}(\vn \ppls \vmu).  
\eqend
It is important to note that due to Proposition 3, there is actually only 
a {\em finite} number of non-zero terms in this sum, i.e.\ 
$\Psi\in\cD_b$. 
With Eq.\ \Ref{c2} the eigenvalue equation $\cH^{\nu,3}\psi = 
\cE\Psi$ implies 
$$
[\cE - \cE_{\nu,N}(\vn\ppls  \vmu ) ]\alpha(\vmu) = \gamma\sum_{\bar 
n=0}^\infty\sum_{j<\ell} \bar n \alpha(\vmu-\bar n \vE_{j\ell} ) 
$$
where 
\eq
\alpha(\vmu):=0\quad  \mbox{ if  $\;\eta_{\nu,N}(\vn \ppls \vmu)=0$} . 
\eqend
Setting $\vmu={ \bf 0 }$ we get, 
\eq
\label{3}
\cE = \cE_{\nu,N}(\vn), 
\eqend 
and $\alpha({\bf 0})$ is arbitrary.  Moreover, 
the other coefficients $\alpha(\vmu)$ are then uniquely determined 
by Eq.\ \Ref{recurr} provided that $b_{\nu,N}(\vn,\vmu):=[
\cE_{\nu,N}(\vn\ppls  \vmu)-\cE_{\nu,N}(\vn ) ]$ 
remains always non-zero. A simple computation shows that
\eq
\label{bnuN}
b_{\nu,N}(\vn,\vmu)=
\left(\frac{2\pi}{L}\right)^2 \sum_{j=1}^N \left( 2\sum_{\ell=j+1}^N 
\mu_{j\ell} \left[n_j-n_\ell + (\ell-j)\nu^2\right] + 
\left[\sum_{\ell=1}^{j-1}\mu_{\ell j} 
- \sum_{\ell=j+1}^N \mu_{j\ell} \right]^2 \right) 
\eqend
which is strictly positive for all $\vmu\in\N_0^{N(N-1)/2}$ if 
\eq
\label{ccc}
n_1\geq n_{2} \geq \cdots \geq n_N  \geq 0
\eqend
(the last inequality here is due to Eq.\ \Ref{cond}). 
Note that Eq.\ \Ref{recurr} then allows to compute the $\alpha(\vmu)$ 
recursively,
\eq
\label{recurr}
\alpha(\vmu) = - \frac{ \gamma}{b_{\nu,N}(\vn,\vmu)} 
\sum_{\bar n=0}^\infty\sum_{j<\ell} 
\bar n \alpha(\vmu-\bar n \vE_{j\ell} ) . 
\eqend
We summarize these calculations and their implication from 
Proposition 4 in the following \\

{\bf \noindent  Theorem 3:} {\em  For $\vn\in\N^N$ satisfying Eq.\ 
\Ref{ccc}, the equations \Ref{notation}--\Ref{recurr} and the normalization 
condition
\eq
\alpha({\bf 0 })=1
\eqend
determine a unique vector $\Psi=\Psi_{\nu,N}(\vn)\in\cD_b$ which is an 
eigenvector of the operator $\cH^{\nu,3}$ with the eigenvalue 
$\cE_{\nu,N}(\vn)$.  Thus 
\eq
\tilde\psi_{\nu,N}(\vn|x_1,\ldots,x_N) :=\lim_{\eps\downarrow 0} 
\left<\Psi_{\nu,N}(\vn) , 
\phi^\nu_\eps(x_1)\cdots \phi^\nu_\eps(x_N) \Omega\right>
\eqend
is in $L^2(S_L^N)$ and is an eigenvector of the CS Hamiltonian 
Eq.\ \Ref{Sutherland} with the same eigenvalue, 
\eq 
\label{Thm3}
[H_{N,\nu^2}- \cE_{\nu,N}(\vn ) ]
\tilde\psi_{\nu,N}(\vn | x_1,\ldots,x_N) = 0 .  
\eqend
}
\vspace*{0.45cm}

{\em Remark:}
We have shown that condition \Ref{ccc} is sufficient for the 
construction of the vectors $\Psi_{\nu,N}(\vn)$ and we show below that 
all eigenvectors of the Sutherland model are thereby obtained.  
Nevertheless, we believe that it would be interesting to explore the 
significance of condition \Ref{ccc} more fully.  However, this is 
beyond the scope of the present paper.  \\

Below we shall compare the eigenfunctions we have obtained with the 
known ones from the literature \cite{Su,Fo2}.  For that it is useful to 
have the corresponding (but simpler) relation for $s=2$, 
\eq
\label{c12}
\cH^{\nu,2}\eta_{\nu,N}(\vn) = \sum_{j=1}^m P_{j} \eta_{\nu,N}(\vn) 
\eqend with $P_{j}$ given in Eq.\ \Ref{Pj}.  This relation is a simple 
consequence of \nonueqa \cH^{\nu,2}\eta_{\nu,N}(\vn) = \sum_{j=1}^N 
\lim_{\eps_1,\ldots,\eps_N\downarrow 0}\int_{-L/2}^{L/2} dy_1 
e^{iP_1y_1} \cdots \int_{-L/2}^{L/2} dy_N e^{iP_N y_N}\nonu\times 
\left(i \frac{\partial}{\partial y_j } \right) 
\phi^{\nu}_{\eps_1}(y_1) \cdots \phi^{\nu}_{\eps_N}(y_N) \Omega 
\nonueqaend which is obtained by using Eqs.\ \Ref{eta}, \Ref{phinup} 
and \Ref{that12}.  By partial integrations we obtain Eq.\ \Ref{c12}.  
Finally using this and an argument similar to that proving 
Theorem 3 we obtain 
\eq
\label{Pjj}
\left( \sum_{j=1}^N i\frac{\partial}{\partial x_j}
- \sum_{j=1}^N P_{j} \right) 
\tilde\psi_{\nu,N}(\vn|x_1,\ldots,x_N)=0 . 
\eqend

\subsection{Relation to Jack polynomials}
We show below that the eigenfunctions of the CS Hamiltonian which we 
have obtained are related to the standard ones in the literature 
\cite{Su} via the following transformation,\footnote{Note that the 
phase factor here is not periodic, thus the wave functions $\psi$ and 
$\tilde\psi$ correspond to different self-adjoint extensions of the 
symmetric operator defined in Eq.\ \Ref{Sutherland}.} \eq
\label{trafo}
\psi_{\nu,N}(\vn,\mu|\vx):= e^{i\pi(\nu^2+2\mu)(x_1+\ldots +x_N)N/L 
}\tilde \psi_{\nu,N}(\vn|\vx) ,\quad \mu\in\N_0 .  \eqend Note that 
the physical interpretation of the phase factor is that it represents 
a free motion (i.e.\ plane wave) of the center-of-mass of the system, 
thus Eq.\ \Ref{trafo} can be regarded as a trivial change of our wave 
functions.

We obtain from Eq.\ \Ref{Thm3} ($p:=\nu^2+2\mu$)
$$
H_{N,\nu^2}\psi_{\nu,N} = e^{i\pi p(x_1+\ldots 
+x_N)N/L }\left( H_{N,\nu^2} - 2\frac{\pi}{L} p N  \sum_{j=1}^N 
i\frac{\partial}{\partial x_j}  + N\left( 
\frac{\pi}{L}p N\right)^2  \right)\psi_{\nu,N},   
$$
and with Eq.\ \Ref{Pjj},
\eq
\label{J1}
H_{N,\nu^2}\psi_{\nu,N}= E\psi_{\nu,N}
\eqend
where
\eq
\label{Ee}
E = \sum_{j=1}^N \left( P_{j}^2 - 2\frac{\pi}{L}p N P_{j} + \left( 
\frac{\pi}{L}p N\right)^2 \right) =
\left(\frac{2\pi}{L}\right)^2\sum_{j=1}^N \left[n_j -\mu 
+ \half\nu^2(N+1-2j)\right]^2  .
\eqend
We thus reproduce all known eigenvalues of the CS Hamiltonian \cite{Su}. 
Note that according to Proposition 2, these eigenfunctions have the form
\eqa
\label{J3}
\psi_{\nu,N}(\vn,\mu| x_1,\ldots,x_N) = e^{2\pi i \mu(x_1+\ldots 
+x_N)N/L } \Delta_{N}^{\nu^2}(x_1,\ldots,x_N) \nonu\times 
\cP_{\vn;\nu}( e^{-2\pi i x_1/L}, \ldots , e^{-2\pi i x_N/L} ) \eqaend 
with $\cP_{\vn;\nu}:=\cP_{\Psi_{\nu,N}(\vn)}$ a symmetric polynomial 
\cite{McD}.  Similarly as in \cite{Fo2} one may use results from 
\cite{St} to prove that the polynomials $\cP_{\vn;\nu}$ are 
proportional to the Jack polynomial associated with the partition 
$\vn$ and the parameter $1/\nu^2$ \cite{St}.  (See Appendix E for 
details.)  It is worth noting that due to this, Theorem 3 can be used 
to formulate an algorithm for explicitly constructing the Jack 
polynomials in terms of the polynomials given in Eq.\ \Ref{polynom}.  
It would be interesting make this algorithm more explicit, but this is 
beyond the scope of the present paper.

\subsection{Duality}
It is known that the eigenfunctions of the CS Hamiltonians Eq.\ 
\Ref{Sutherland} with couplings $\beta=\nu^2$ and $\beta=1/\nu^2$ are 
closely related to each other \cite{St}.  In our approach this duality 
appears as follows.

We note that Eqs.\ \Ref{Hpmnu} implies
\eq
\cH^{\nu,3}= -\nu^2 \cH^{-1/\nu,3} +\nu[ W^{\nu,3} - 
W^{-1/\nu,3} ] .  
\eqend
Using Eq.\ \Ref{res21} we obtain by a straightforward computation
\eqa
\nu[ W^{\nu,3} - 
W^{-1/\nu,3} ] = 
-4\frac{\pi}{L}(\nu^2+1) Q W^{-1/\nu,2} +E_0(Q;\nu)
\eqaend
with 
\eq
\label{E0}
E_0(Q;\nu)=c_{\nu,L} \left( 
4(4\nu^4-3\nu^3-4\nu^2-9\nu-5)Q^2 - 3\nu^4+2\nu^3+5\nu^2-2\nu-3) 
\right) Q
\eqend 
where $c_{\nu,L}= \frac{2(\nu^2+1)}{3\nu^2}\left(\frac{\pi}{L}\right)^2$.
By an argument similar to the one leading to Theorem 3 we conclude: 

{\bf \noindent  Theorem 4:} {\em  The vectors 
$\Psi_{-1/\nu,N}(\vn)\in\cD_b$ characterized in Theorem 3 are
eigenvectors of the operator $\cH^{\nu,3}$ with corresponding 
eigenvalues
\eqa
\tilde\cE_{\nu,N}(\vn )= 
-\nu^2 \sum_{j=1}^N \tilde P_j^2 -4\frac{\pi}{L}(\nu^2+1)N  \sum_{j=1}^N 
\tilde P_j + \tilde E_0(N,\nu)
\eqaend
where $\tilde P_j:=P_{j,-1/\nu,N}$ as defined in Eq.\ 
\Ref{Pj} and $E_0(N,\nu)$ Eq.\ \Ref{E0}. 
Thus 
\eq
\lim_{\eps\downarrow 0} \left<\Psi_{-1/\nu,N}(\vn) , 
\phi^\nu_\eps(x_1)\cdots \phi^\nu_\eps(x_N) \Omega\right>
\eqend
is an eigenvector of the CS Hamiltonian $H_{N,\nu^2}$ 
Eq.\ \Ref{Sutherland} with the same eigenvalue 
$\tilde\cE_{\nu,N}(\vn )$. }  

\vspace*{0.45cm}

\subsection*{Appendix E. On Jack polynomials}
As discussed in Ref.\ \cite{Fo2} (see also \cite{Su}), all 
eigenvalues of the CS Hamiltonian $H_{N,\nu^2}$ are of the form
Eq.\ \Ref{Ee} 
with $\mu$ a non-positive integer and the  $n_j$ integers such that 
$n_1\geq n_2\geq \ldots n_N\geq 0$. Moreover, the corresponding 
eigenfunctions are given by\footnote{This is Eq.\  
(2.16) in \cite{Fo2} with $\mu_N$ replaced by $-\mu$.}
\eq
\label{ee}
\psi = e^{-2\pi i N\mu(x_1+\ldots+x_N)/L } C_{\vn}^{(1/\nu^2)}(e^{2\pi 
i x_1/L},\ldots,e^{2\pi i x_1/L}) \Delta^{\nu^2}(x_1,\ldots,x_N) 
\eqend where $C_{\vn}^{(\alpha)}$ is the Jack polynomial \cite{St} 
associated with the partition $\vn$ and parameter $\alpha$ \cite{Fo2}.  
Note that the complex conjugate $\psi^*$ of $\psi$ is also an 
eigenfunction of $H_{N,\nu^2}$ with the same eigenvalue, and our 
eigenfunction Eq.\ \Ref{J3} has the same form as $\psi^*$.  Note also 
that $\psi^*$ can be written also in the form Eq.\ \Ref{ee} with the 
parameters $\mu,n_j$ replaced by $\mu',n_j'$ which are such that \eq 
n_j'-\mu' = \mu - n_{N-j} ,\quad n_N'\geq 0.  \eqend This follows from 
the fact that $E$ Eq.\ \Ref{Ee} is invariant under the transformation 
$\mu,n_j \to \mu',n_j'$.

We now derive the precise relation of our solutions to the Jack 
polynomials.  Similarly to \cite{Fo2} we 
deduce from Eqs.\ \Ref{J1}--\Ref{J3} that 
$\cP_{\vn;\nu}=\cP_{\vn;\nu}(e^{-2\pi i x_1/L},\ldots,e^{-2\pi i 
x_1/L})$ obeys the equation\footnote{This is Eq.\ (2.3) in \cite{Fo2} 
adapted to our notation.  Note that $\gamma$ in \cite{Fo2} corresponds 
to $2\nu^2$ here.} \eq -\sum_{j=1}^N\frac{\partial^2 \cP_{\vn;\nu} 
}{\partial x_j^2} - \frac{2\pi \nu^2}{L}\sum_{1\leq j<k\leq 
N}\cot\frac{\pi(x_k-x_j)}{L} \left( \frac{\partial}{\partial z_k} 
-\frac{\partial}{\partial z_j} \right)\cP_{\vn;\nu} = 
(E-E_0)\cP_{\vn;\nu} \eqend where
\eq E-E_0 = \left(\frac{2\pi}{L}\right)^2 
\sum_{j=1}^N \left( n_j ^2 + \nu^2 n_j ( N + 1 - 2j ) \right) .  
\eqend Moreover, \eq \sum_{j=1}^N\frac{\partial \cP_{\vn;\nu} 
}{\partial x_j} = -i\frac{2\pi}{L}\sum_{j=1^N} n_j \cP_{\vn;\nu} 
\eqend follows from Eq.\ \Ref{Pjj}.  Comparing with the differential 
equation defining the Jack polynomials \cite{St}, these equations 
imply that $\cP_{\vn;\nu}(e^{-2\pi i x_1/L},\ldots,e^{-2\pi i x_1/L})$ 
equals, up to a constant, the Jack polynomial 
$C_{\vn}^{(1/\nu^2)}(e^{-2\pi i x_1/L},\ldots,e^{-2\pi i x_1/L})$ 
associated with the partition $\vn$ \cite{Fo2}.

\end{document}